\documentclass[letterpaper,11pt]{article}
\usepackage{caption}
\pdfoutput=1 % if your are submitting a pdflatex (i.e. if you have
\usepackage{extarrows}
\usepackage{amssymb}
\usepackage{amsmath}

\usepackage{amstext}
\usepackage{etoolbox}
\let\tinymatrix\smallmatrix

\patchcmd{\tinymatrix}{\scriptstyle}{\scriptscriptstyle}{}{}
\patchcmd{\tinymatrix}{\scriptstyle}{\scriptscriptstyle}{}{}
\patchcmd{\tinymatrix}{\vcenter}{\vtop}{}{}
\patchcmd{\tinymatrix}{\bgroup}{\bgroup\scriptsize}{}{}

\usepackage{graphicx,epsfig}
\usepackage{epsfig}
\usepackage{verbatim} 
\usepackage{color}
\usepackage{ulem}
\usepackage{enumitem}
\usepackage{subfigure}
\usepackage{bbm}
\usepackage{parskip}
\usepackage{dsfont}
\usepackage[numbers,sort&compress]{natbib}
\usepackage[body={6.5in, 9in}, right=1in, top=1in]{geometry}
\usepackage{mathtools}
\usepackage{comment}
\usepackage{caption}
\usepackage{bbold}
\usepackage{float}

\newcommand{\Comment}[1]{{}}
\definecolor{darkblue}{rgb}{0.15,0.35,0.55}
\definecolor{reddish}{rgb}{0.65, 0.2, 0.2}
\usepackage[linktocpage=true]{hyperref}
\newcommand{\be}{\begin{equation}}
\newcommand{\ee}{\end{equation}}
\newcommand{\bea}{\begin{eqnarray}}
\newcommand{\eea}{\end{eqnarray}}
\newcommand{\beas}{\begin{eqnarray*}}
\newcommand{\eeas}{\end{eqnarray*}}

\def\({\left(}
\def\){\right)}

\def\gsim{ \lower .75ex \hbox{$\sim$} \llap{\raise .27ex \hbox{$>$}} }
\def\lsim{ \lower .75ex \hbox{$\sim$} \llap{\raise .27ex \hbox{$<$}} }

\usepackage{xcolor,colortbl}

\begin{document}
\def\thefootnote{\fnsymbol{footnote}}

\begin{center}
\LARGE{\textbf{Accessibility Measure for Eternal Inflation:\\Dynamical Criticality and Higgs Metastability}} \\[0.5cm]
 
\large{Justin Khoury}
\\[0.5cm]

\small{
\textit{Center for Particle Cosmology, Department of Physics and Astronomy, University of Pennsylvania,\\ 209 South 33rd St, Philadelphia, PA 19104}}

\vspace{.2cm}

\end{center}

\vspace{.6cm}

\hrule \vspace{0.2cm}
\centerline{\small{\bf Abstract}}
%\vspace{-0.2cm}
{\small\noindent We propose a new measure for eternal inflation, based on search optimization and first-passage statistics. This work builds on the dynamical selection mechanism for vacua based on search optimization proposed recently by the author and Parrikar. The approach is motivated by the possibility that eternal inflation has unfolded for a finite time much shorter than the exponentially long mixing time for the landscape. The proposed {\it accessibility measure} assigns greater weight to vacua that are accessed efficiently under time evolution. It is the analogue of the closeness centrality index widely used in network science. The proposed measure enjoys a number of desirable properties. It is simultaneously time-reparametrization invariant, independent of initial conditions, and oblivious to physical {\it vs} comoving weighing of pocket universes. Importantly, the proposed measure makes concrete and testable predictions that are largely independent of anthropic reasoning. Firstly, it favors vacua residing in regions of the landscape with funnel-like topography, akin to the energy landscape of naturally-occurring proteins. Secondly, it favors regions of the landscape that are tuned at dynamical criticality, with vacua having an average lifetime of order the de Sitter Page time. Thus the predicted lifetime of our universe is of order its Page time, $\sim 10^{130}$~years, which is compatible with Standard Model estimates for electroweak metastability. Relatedly, the supersymmetry breaking scale should be high, at least $10^{10}$~GeV. The discovery of beyond-the-Standard Model particles at the Large Hadron Collider or future accelerators, including low-scale supersymmetry, would rule out the possibility that our vacuum lies in an optimal region of the landscape. The present framework suggests a correspondence between the near-criticality of our universe and dynamical critical phenomena on the string landscape. 
\vspace{0.3cm}
\noindent
\hrule
\def\thefootnote{\arabic{footnote}}
\setcounter{footnote}{0}

\section{Introduction}

If the fundamental theory allows eternal inflation~\cite{Vilenkin:1983xq,Linde:1986fc,Linde:1986fd}, then a {\it measure} is necessary to assign probabilities for
different outcomes and therefore make predictions. The task of defining a measure satisfying various physical desirata, such as time-reparametrization
invariance and independence on initial conditions, has proven to be formidably challenging. This is the well-known measure problem. (See~\cite{Freivogel:2011eg} for a review.)
In the 80's and 90's the measure problem was viewed by and large as a pesky mathematical subtlety, with the exception of a handful of inflationary cosmologists
who studied it seriously. In the early 2000's, it came to more widely appreciated as a pressing problem with the discovery of an exponentially large number of metastable
de Sitter (dS) vacua in string theory~\cite{Bousso:2000xa,Kachru:2003aw,Susskind:2003kw,Douglas:2003um}.\footnote{Recently, the existence of metastable dS vacua in
string theory, at least in parametrically-controlled regimes, has been questioned through the dS swampland conjecture~\cite{Obied:2018sgi,Agrawal:2018own,Garg:2018reu,Ooguri:2018wrx}.
This has sparked a heated debate --- see~\cite{Palti:2019pca} and references therein. In our analysis, we assume for concreteness the existence of a landscape of metastable dS vacua.}

Historically two broad classes of measures have been considered:

\begin{itemize}

\item {\it Global measures} define a global foliation of space-time specified by a global time coordinate~$t$. One counts pocket universes\footnote{Pocket universes can consist either of thermalized regions, if the evolution on the landscape is dominated by quantum diffusion of scalar fields, or false-vacua dS regions, if the evolution is instead dominated by quantum tunneling and bubble nucleation. To simplify the discussion, in this paper we will focus exclusively on the latter case for concreteness.} on a late-time cutoff surface~$t = t_{\rm c}$, and then lets~$t_{\rm c}\rightarrow \infty$~\cite{Linde:1993nz,Linde:1993xx,GarciaBellido:1993wn,Vilenkin:1994ua,Guth:2007ng,Garriga:1997ef,Garriga:2005av,Vanchurin:2006qp,Bousso:2008hz,DeSimone:2008if,DeSimone:2008bq}. This prescription has the advantage of being independent of initial conditions, consistent with the attractor property of inflation. Its major drawback is that the result depends sensitively on the choice of foliation~\cite{Linde:1993nz,Linde:1993xx,GarciaBellido:1993wn}. Another source of ambiguity is whether pocket universes are weighted according to their comoving or physical volume. This can result in exponentially different results, even for fixed foliation.

Faced with these difficulties, recent work on global measures has focused on ruling out certain foliations based on phenomenological input, {\it e.g.} avoiding the youngness paradox~\cite{Guth:2007ng} or domination by Boltzmann brains~\cite{Bousso:2008hz,DeSimone:2008if}. This has led to some convergence towards scale-factor time as the most phenomenologically sensible foliation~\cite{DeSimone:2008if}.\footnote{Recently the 4-volume cutoff measure has been proposed~\cite{Vilenkin:2019mwc}, which improves on certain technical drawbacks of the scale factor measure while otherwise making similar predictions.} While this interplay between theory and phenomenology is logically reasonable, it would be far more desirable to define {\it ab initio} a measure based solely on theoretical principles, such as time-reparametrization invariance and independence on initial conditions, and derive phenomenological implications {\it a posteriori}.  

\item {\it Local measures} focus on a space-time region around a time-like observer. Examples include the past light-cone of world-lines (causal diamond measure~\cite{Bousso:2006ev,Bousso:2009dm}), a region bounded by the apparent horizon~\cite{Bousso:2010zi}, and a space-like region around a world-line ({\it e.g.}, the ``watcher" measures~\cite{Garriga:2005av,Vanchurin:2006qp,Garriga:2012bc,Nomura:2011dt}). Such constructions are manifestly gauge-invariant. However, because a typical geodesic will eventually enter an AdS or Minkowski vacuum, generally assumed to be terminal,\footnote{An alternative and more speculative possibility is that collapsing AdS regions can sometimes bounce and avoid big crunch singularities~\cite{Garriga:2012bc,Garriga:2013cix}. In this paper we will treat AdS regions as terminal.} all but a measure zero of watchers will sample a finite number of bubbles. This regulates the infinities of eternal inflation~\cite{Bousso:2006ev}, at the expense of sensitivity to initial conditions.\footnote{A possible argument~\cite{Bousso:2006ev,Bousso:2009dm} is that the question of initial conditions is logically distinct from the measure problem and should be provided by the theory of quantum gravity. While this is a fair point, it would be more satisfactory to have a measure, defined solely within semi-classical gravity, that is both time-reparametrization invariant {\it and} independent of initial conditions.} In particular, it has been shown that for specific choices of initial conditions certain local measures agree with their global counterpart~\cite{Bousso:2009dm,Bousso:2009mw}.

\end{itemize} 

Despite the variety of approaches, all proposed measures to date focus on the same statistics: {\it the stationary distribution of a Markov
process describing vacuum dynamics.} Mathematically, the landscape can be modeled by a graph or network whose nodes represent the different vacua,
and whose links denote the relevant transitions (Sec.~\ref{RW complex nets}). After coarse-graining, the probability $f_i(t)$ that a time-like observer
(``watcher") occupies vacuum $i$ at time $t$ is governed by a linear Markov equation~\cite{Garriga:1997ef,Garriga:2005av}. (Equivalently, $f_i(t)$
is the fraction of comoving volume occupied by vacuum $i$.) 

For generic initial conditions, the solution to the Markov process tends asymptotically to a stationary distribution: $f(t)\rightarrow f^{\infty}$. This stationary distribution is the zero-mode of the transition matrix, and as such lies entirely within the subspace of terminal vacua. The relative probability to lie in different non-terminal (dS) vacua is therefore determined by the subleading term as $t\rightarrow \infty$, which is in turn set by the slowest-decaying (or ``dominant") eigenvector: $\delta f \sim v^{(1)}_{M} {\rm e}^{\lambda_1 t}$. Here $\lambda_1 < 0$ is the ``dominant" eigenvalue, {\it i.e.}, the non-zero eigenvalue of~$M$ with the smallest magnitude, which sets the relaxation time. In turn,~$v^{(1)}_{M}$ is dominated by the dS vacuum with the {\it slowest decay rate} anywhere on the landscape. Since this so-called ``dominant vacuum" is unlikely to be hospitable, the relative probabilities for different hospitable vacua are determined by the transition rate from the dominant vacuum to each hospitable vacuum. All global and local measures are based on this late-time prescription. 

Aside from the technical challenges mentioned earlier, the late-time prescription presents another drawback --- its sensitivity to exponentially small terms in the transition matrix. The argument, laid out in the Appendix, can be summarized as follows. Although the detailed nature of the dominant vacuum requires input from string theory, one expects on general grounds that it has very small vacuum energy $V_{\rm dom}$, and is surrounded by vacua of much higher potential energy. In this configuration, its only allowed Coleman-De Luccia (CDL) transitions involve ``up-tunneling". By detail balance, the rate for such upward transitions is exponentially suppressed by ${\rm e}^{-48\pi^2M_{\rm Pl}^4/V_{\rm dom}}$ compared to the (already exponentially small) rate for the reverse processes. In particular, transition rates from the dominant vacuum to hospitable vacua, which set the relative probabilities, are sensitive to exponentially small contributions to the transition matrix. Minor tweaks to the landscape can alter these exponentially small corrections, resulting in different predictions. While not logistically inconsistent, we view such sensitivity to minor tweaks to the landscape as undesirable.

\subsection{A new measure of centrality}

In the language of graph theory the problem of defining a measure for eternal inflation is a question of {\it centrality} --- which nodes in the network are,
in a suitably defined sense, most important? Recently there has been tremendous activity in understanding the properties of real-world networks,
such as the world wide web, academic co-authorship, social networks, protein interaction networks, epidemic propagation {\it etc.} The question of
centrality is of utmost importance in these studies. 

Various centrality indices have been proposed, and they can be broadly classified as followed (see~\cite{centrality review} for a review). 
One category of centrality measures is based on spectral properties of the transition matrix. This includes eigenvector centrality, Katz index~\cite{Katz}, and
Google's PageRank algorithm~\cite{pagerank}. The stationary distribution on which global and local measures are based belongs to this category.
Specifically it is an example of left dominant eigenvector centrality. 

Another category of centrality measures focuses instead on the shortest paths between nodes. For instance, the {\it closeness}
measure~\cite{closeness1,closeness2} assigns greater weight to nodes that can be reached on average with the fewest number of steps.
Another example is {\it betweenness centrality}, which favors nodes that have the most shortest paths between other pairs of nodes passing through them.
Intuitively, nodes that are favored by closeness or betweenness centrality are most important in controlling the flow of information in the network.

In this paper we present a new measure for eternal inflation that is analogous to closeness centrality. We call it
the {\it accessibility measure}. Instead of characterizing the distribution of vacua at equilibrium, the proposed
measure pertains to the {\it approach to equilibrium}. Indeed, since eternal inflation is past-geodesically incomplete~\cite{Borde:2001nh},
eternal inflation started a finite proper time in our past. The standard approach based on the stationary distribution assumes
that we live sufficiently long after the onset of eternal inflation, such that vacuum statistics have reached equilibrium. 
While logically consistent, this assumption is non-trivial. Globally the landscape features many exponentially long-lived
metastable vacua, resulting in glassy dynamics and exponentially long relaxation time~\cite{Denef:2011ee}.  

The accessibility measure is motivated instead by the alternative possibility that eternal inflation has unfolded for a time
much shorter than the relaxation time. In this case, as first proposed in~\cite{Denef:2017cxt}, a vacuum like ours should be likely not
because it is typical according to the stationary distribution, but rather because it has the right properties to be accessed early
on in the evolution. At times much smaller than the mixing time, most hospitable vacua have been accessed once or perhaps not at all.
In such a situation, the occupational probability $f_i(t)$ is not the most reliable statistics. Instead the accessibility measure will be based
on {\it first-passage statistics}, which are best-suited to study the approach to equilibrium. The accessibility measure will give greater weight to
vacua with high first-passage probability, or equivalently short first-passage time. Such vacua are most likely to be accessed quickly, much earlier than the relaxation time. 

A natural concern with working at early times is sensitivity to initial conditions. The accessibility measure will be defined by effectively averaging over initial conditions,
to give a result that is independent of initial conditions as well as time-reparametrization invariant. Furthermore, because it is based on access time instead of volume fractions, it is oblivious to
comoving {\it vs} physical volume ambiguity and avoids any youngness bias. The accessibility measure is also insensitive to minor tweaks to the landscape; it can be reliably calculated by
neglecting the exponentially small terms in the transition matrix that encode upward-tunneling. Last but not least, the accessibility measure makes testable and falsifiable predictions (summarized below).

\begin{figure}[htb]
\centering
\includegraphics[height=2.5in]{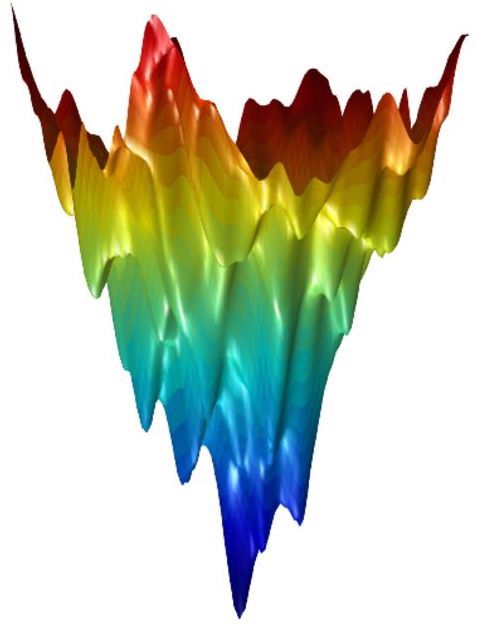}
\caption{An optimal region is characterized by all vacua having at least one allowed downward transition.
Its landscape topography is that of a funnel, akin to the free energy landscape of proteins. (Reproduced from~\cite{funnelfig}.)}
\label{optimal region}
\end{figure}

This work is a continuation of a recent paper~\cite{Khoury:2019yoo}, which presented a new dynamical selection mechanism for
vacua based on search optimization. Instead of restricting to late-time, stationary distributions for the different vacua, 
the analysis focused on the approach to equilibrium. It was shown that the mean first-passage time (MFPT) to hospitable vacua is minimized for vacua lying at the
bottom of funnel-like regions of the landscape, as sketched in Fig.~\ref{optimal region}. This is akin to the smooth folding funnels of
naturally-occurring proteins~\cite{proteins1,proteins2}, where the high-energy unfolded states are connected to the lowest-energy native state
by a relatively smooth funnel. Furthermore, it was argued that optimal regions of the landscape achieve a compromise between
minimal oversampling and sweeping exploration of the region, resulting in dynamical criticality. 

The analysis of~\cite{Khoury:2019yoo} stopped short, however, of defining an actual measure for the likelihood of residing in
optimal regions of the landscape. The goal of this paper is to fill this gap. We will argue that the accessibility measure 
favors vacua in optimal regions, characterized by funnel-like topography and dynamical criticality. 

\subsection{Brief overview of results}

Our measure is simplest to define in a toy landscape comprised only of dS vacua (Sec.~\ref{measure dS}).
The key building block is the MFPT $\langle t_{j\rightarrow i}\rangle$ between a pair of vacua~$i$ and~$j$, defined as the time for a random walker starting from~$j$ to reach~$i$ averaged
over all paths connecting the two nodes~\cite{MFPT book}. We then define a dimensionless {\it partial MFPT} (pMFPT) to a given node~$i$
by averaging the MFPT over all initial nodes~$j$
\be
{\cal T}_i \equiv \frac{v^{(1)\,2}_i }{1+v^{(1)\,2}_i}\sum_{j\neq i} v^{(1)\,2}_j  \frac{\langle t_{j\rightarrow i}\rangle}{\Delta t} \,,
\label{GMFPTintro}
\ee 
where $\Delta t$ is the coarse-graining time step, and~$v^{(1)}_j$ is the zero-mode of the transition matrix which sets the stationary distribution
for the occupational probability.\footnote{The partial MFPT~\eqref{GMFPTintro} is closely related to a quantity introduced by~\cite{GMFPT}, 
which was called global MFPT.} Therefore~${\cal T}_i$ can be thought of as the MFPT to node~$i$, averaged over all initial nodes weighted
by the stationary distribution. The overall factor of~$\frac{v^{(1)\,2}_i }{1+v^{(1)\,2}_i}$ is included to simplify some of the results below.

At first sight it may seem strange to weigh initial vacua according to the stationary distribution, since our measure is supposed
to capture information about the approach to equilibrium. We offer the following justification. Given a Markov process, one
can argue that the stationary distribution is a natural distribution to consider.\footnote{Incidentally, it is by similarly averaging over initial conditions using
the stationary distribution that certain local measures match the predictions of their global counterpart~\cite{Bousso:2009dm,Bousso:2009mw}.} More to the point,
however, the stationary distribution offers a {\it conservative} estimate of the average time needed to reach a given node. Indeed, as we will see,~$v^{(1)\,2}_j$
is dominated by the most stable dS vacuum, which should therefore correspond to the largest MFPT~$\langle t_{j\rightarrow i}\rangle$.
In other words, the weighted sum in~${\cal T}_i$ gives greatest weight to the initial node corresponding to the longest average
travel time to~$i$. From this point of view we expect that any distribution with non-zero support on the most stable dS vacuum (such as a uniform distribution)
should give similar results. That said, we cannot claim that the accessibility measure defined below is unique.

The pMFPT can be expressed in alternative ways, each offering different insights. Firstly, we will see that~\eqref{GMFPTintro}
can be cast simply in terms of the eigenvalues and eigenvectors of the transition matrix. This makes the
time-reparametrization invariance of~${\cal T}_i$ manifest.  Secondly, it can be written in terms of first-return statistics~\cite{GMFPT}:
\be
{\cal T}_i  = \frac{1}{2} \frac{\langle t_{i\rightarrow i}^{\,2}\rangle}{\langle t_{i\rightarrow i}\rangle^2}\,,
\label{GMFPTintro2}
\ee
where $\langle t_{i\rightarrow i}\rangle$ is the first-return time, and $\langle t_{i\rightarrow i}^{\,2}\rangle$ is the
second moment of the first-return probability. Since by definition first-return statistics for~$i$ describe random walks
that all start at~$i$, this expression makes manifest that~${\cal T}_i$ is independent of initial conditions. Furthermore,
it implies an important lower-bound~\cite{GMFPT}:
\be
{\cal T}_i  \geq \frac{1}{2}\,.
\label{Tjbound}
\ee

A third, and perhaps most intuitive, interpretation of the pMFPT is in terms of first-passage probability. On a finite landscape
every vacuum is guaranteed to be populated eventually, but the required time scale for {\it all} vacua to be accessed is of course
the relaxation time. At times much shorter than the relaxation time, we will see that maximizing the first-passage probability to a given vacuum
is equivalent to minimizing its pMFPT. 

Yet another equivalent expression for the pMFPT is in terms of the escape or never-return probability~$S_i$ from~$i$ in the limit of an infinite network.
This represents the probability that a random walker who started from~$i$  eventually never returns to $i$. We will find the simple relation ${\cal T}_i = S_i^{-1}$. 

Our proposed accessibility measure for dS-only landscapes is then defined as\footnote{The accessibility measure defined here is related to random walk centrality
introduced in~\cite{complex network PRL}, as well as second order centrality studied in~\cite{2ndorder}.}
\be
p_i = \frac{{\cal T}_i^{-1}}{\sum_k {\cal T}_k^{-1}} \,.
\label{pj}
\ee
By construction, the measure is both time-reparametrization and independent of initial conditions. Furthermore, it is oblivious
to the comoving {\it vs} physical volume ambiguity. The measure favors vacua that are easily
accessed under time evolution, {\it i.e.}, those vacua that saturate~\eqref{Tjbound}: ${\cal T}_i \sim {\cal O}(1)$.
Thus, unlike standard measures based on the stationary distribution, the proposed measure does
not exponentially favor a single vacuum. There can be many vacua with~${\cal T}_i \sim {\cal O}(1)$,
and all will be weighted equally according to~\eqref{pj}. 

In Sec.~\ref{measure with terms} we will generalize the accessibility measure to a landscape with terminal vacua.
For dS vacua, we have mentioned above four different equivalent expressions for the pMFPT in the case
of a dS-only toy landscape: 1.~In terms of a weighted average of the MFPT to a given node; 2.~In terms
of the variance of first-return times to a given node; 3. In terms of the first-passage probability for times shorter than the
relaxation time; 4.~In terms of the escape probability from a given node.

Once terminals are included, however, these give inequivalent definitions of the pMFPT.  It turns out that the last definition in terms of the escape probability offers the most straightforward generalization
to the case with terminals. Therefore for dS vacua we define the accessibility measure in terms of the
escape probability. For terminal vacua, the natural analogue to an escape probability is a trapping probability.
Both escape and trapping probabilities admit simple expressions in terms of eigenvectors and eigenvalues
of the transition matrix. The resulting accessibility measure is both time-reparametrization invariant and independent
of initial conditions.

\subsection{Predictions of the accessibility measure}
\label{predictions intro}

Importantly, the accessibility measure makes concrete and testable predictions (Sec.~\ref{pheno}), in particular for new
physics (or absence thereof) at the Large Hadron Collider (LHC). These predictions are largely independent of anthropic
reasoning. The only input of possible anthropic origin is the vacuum energy, or equivalently the Hubble constant~$H_0$.
Given the observed value of~$H_0$, we derive predictions for other observables, such as the optimal lifetime of our universe,
which follow readily from the measure itself.   

\begin{itemize}

\item {\bf Access time:} The accessibility measure favors vacua that are easily accessed under time evolution.
Specifically, it favors vacua that nearly saturate~\eqref{Tjbound}, {\it i.e.}, with order unity pMFPT. In terms of proper time for
vacuum~$I$, this implies an optimal access time of order its Hubble time: 
\be
\tau^{\rm access}_I \sim H_I^{-1}\,.
\label{access time Hubble time intro}
\ee
Given the observed vacuum energy in our universe, this implies that we live approximately $H_0^{-1} \sim 13.8$~billion
years after the beginning of eternal inflation. This in turn implies an upper bound on the number of e-folds during
the last period of inflation prior to reheating: ${\cal N} \;\lsim\; H_{\rm inf}/H_0$.

\item {\bf Funnel topography:} To make further predictions we will consider a finite region of the
landscape that includes~$N_{\rm inf}\gg 1$ dS vacua. The fiducial region is treated as a closed system for simplicity, though
this assumption is relaxed later on. We define in Sec.~\ref{funnel sec} a characteristic time~$\langle {\cal T} \rangle$ for
the landscape dynamics of dS vacua in the region, as a suitable average over the pMFPTs. In the downward
approximation, where upward transitions are neglected, this average pMFPT
reduces to 
\be
\langle {\cal T} \rangle \simeq \frac{1}{N_{\rm inf}} \sum_{i=1}^{N_{\rm inf}} \frac{1}{\kappa_i \Delta t}\,,
\label{avg T down intro}
\ee
where $\kappa_i$ is the total decay rate of the $i^{\rm th}$ vacuum. Thus~$\langle {\cal T} \rangle$ is recognized as a {\it mean residency time}.

If any vacuum in the region has only upward-tunneling as allowed transitions, then its decay rate will be exponentially small, resulting in an
exponentially large~$\langle {\cal T} \rangle$. Such a region is characterized by frustration and glassy dynamics~\cite{glassylandscape}. It
should be clear that such frustrated regions are heavily disfavored by the accessibility measure. Instead, the accessibility measure
favors regions whose dS vacua all have allowed downward transitions, either to lower-lying dS vacua or to terminals. Such favored regions
therefore have the topography of a {\it broad valley or funnel}, as shown in Fig.~\ref{optimal region}. This reaffirms the result
obtained in~\cite{Khoury:2019yoo}, this time motivated by a well-defined measure. 

This is a key prediction of the accessibility measure. Unlike stationary measures, according to which our vacuum should be generic
among all hospitable vacua on the landscape (the ``principle of mediocrity"), {\it the accessibility measure favors vacua residing in special regions 
of the landscape with funnel-like topography.}

\item {\bf Dynamical criticality and Page lifetime:} To properly treat regions of the landscape as open systems
would require modeling their environment. Following~\cite{Khoury:2019yoo}, we instead study a proxy requirement that relies
solely on the intrinsic dynamics within a given region. Specifically, we demand that random walks in the region are {\it recurrent}
in the infinite-network limit. Recurrent walks thoroughly explore any region around their starting point,
hence recurrence offers a reliable and model-independent benchmark for efficient sampling~\cite{Khoury:2019yoo}.

We will show in Sec.~\ref{criticality page} that the joint demands of minimal oversampling, defined by minimal~$\langle {\cal T} \rangle$,
and sweeping exploration, defined by recurrence, selects regions of the landscape that are tuned at {\it dynamical criticality}.
Vacua in optimal regions have an average lifetime of order the dS Page time:
\be
\tau_{\rm crit} (H) \sim \frac{M_{\rm Pl}^2}{H^3}\,.
\label{tau crit intro}
\ee
Thus the Page time represents an optimal time for vacuum selection, as first realized in~\cite{Khoury:2019yoo}.

\item {\bf Higgs metastability and particle phenomenology:} For our vacuum,~\eqref{tau crit intro} implies the lifetime
\be
\tau_{\rm decay} \sim \frac{M_{\rm Pl}^2}{H_0^3} \sim 10^{130}~{\rm years}\,.
\label{tdecay pred intro}
\ee
Remarkably, this agrees to within~$\gsim\; 2\sigma$ with the Standard Model (SM) estimate for electroweak
metastability~\cite{Andreassen:2017rzq}:~$\tau_{\rm SM} = 10^{526^{+409}_{-202}}~{\rm years}$. In other words, taking $H_0$ as input,
the optimal lifetime~\eqref{tdecay pred intro} constrains a combination of SM parameters, in particular the Higgs and top quark masses. Closer
agreement with the SM lifetime estimate can be achieved if the top quark is slightly heavier, $m_{\rm t} \simeq 174.5~{\rm GeV}$, or with
new physics at intermediate scales, such as right-handed neutrinos with mass of $10^{13}-10^{14}~{\rm GeV}$~\cite{EliasMiro:2011aa}.

More generally, the accessibility measure offers a dynamical explanation for the near-criticality of our vacuum. It gives a {\it raison d'\^{e}tre}
for the conspiracy underlying Higgs metastability. Therefore, from this point of view the inferred metastability of the electroweak vacuum
is sacred. New Beyond-the-SM (BSM) physics discoverable by the LHC, on the other hand, can jeopardize this
observable and, barring fine-tunings, will make our vacuum stable. Therefore, {\it the discovery of BSM particles at the LHC and future colliders, including low-scale SUSY, would rule out the possibility that our vacuum lies in an optimal region of the landscape.} This is a falsifiable prediction of the accessibility measure.

\item {\bf Scale of inflation:} If our vacuum lies in an optimal region, then, on the one hand, it was accessed within a Hubble time~$H_0^{-1}$,
per~\eqref{access time Hubble time intro}, and, on the other hand, originated from a parent vacuum whose lifetime was of order the Page time. These two facts together imply
a bound on the Hubble scale of the parent vacuum: $H_{\rm parent} \;\gsim\; \left(M_{\rm Pl}^2 H_0\right)^{1/3}$. It is usually assumed that the tunneling event from the parent vacuum is followed by a period of slow-roll inflation, with Hubble scale~$H_{\rm inf}$. Assuming that $H_{\rm inf} \sim H_{\rm parent}$, then~\eqref{H parent bound}
implies a lower bound on the slow-roll inflationary energy scale:
\be
E_{\rm inf} \sim \sqrt{H_{\rm parent}M_{\rm Pl}} \;\gsim\; 10^8~{\rm GeV} \,.
\label{Einf lower intro}
\ee
Meanwhile, it has been argued that if the inflationary is too high, then Higgs quantum fluctuations during inflation
could push the field beyond the potential barrier~\cite{Espinosa:2015qea}. Assuming minimal coupling of the Higgs to gravity,
for simplicity, Higgs fluctuations will be under control if the inflationary scale satisfies
\be
E_{\rm inf}\;\lsim\; 10^{14}~{\rm GeV} \,.
\label{Einf upper intro}
\ee 
(The bound becomes looser with non-minimal coupling of suitable sign~\cite{Espinosa:2015qea}. It has also been argued that the bound is also sensitive to higher-dimensional operators and deviations from exact de Sitter~\cite{Fumagalli:2019ohr}.) Equations~\eqref{Einf lower intro} and~\eqref{Einf upper intro} together imply an optimal range for the inflationary scale, assuming a minimally-coupled Higgs, of $10^8~{\rm GeV}\;\lsim\; E_{\rm inf}\;\lsim\; 10^{14}~{\rm GeV}$. Therefore, a detection of primordial gravitational waves, for instance from cosmic microwave background polarization, would either imply that the Higgs must have a non-minimal coupling to gravity, or otherwise disfavor the possibility that our vacuum lies in an optimal region of the landscape. 

\end{itemize}

Our results can also be cast in the language of computational complexity and search optimization. As mentioned earlier, the landscape features many very long-lived false vacua,
resulting in frustrated dynamics and exponentially long mixing time~\cite{Denef:2011ee}. Correspondingly, it has been shown~\cite{Denef:2006ad} in the context of simplified landscape models~\cite{Bousso:2000xa,ArkaniHamed:2005yv} that finding a vacuum within a hospitable range of potential energy is an~\textsf{NP}-hard problem. See also~\cite{Bao:2017thx,Halverson:2018cio}. Aside from computational complexity, it has been argued that string compactifications also face issues of undecidability~\cite{Cvetic:2010ky,Halverson:2019vmd}. There has been much activity recently in applying deep learning algorithms to the string landscape search problem~\cite{Carifio:2017nyb,He:2017aed,Krefl:2017yox,Ruehle:2017mzq,Carifio:2017bov,Wang:2018rkk,Klaewer:2018sfl,Mutter:2018sra,Cole:2018emh,Halverson:2019tkf,He:2019vsj,Cole:2019enn}.

We will show that regions with slow (glassy) transition rates correspond to a mean residency time~$\langle {\cal T} \rangle$ scaling polynomially in~$N_{\rm inf}$.
Since~$N_{\rm inf}$ generically scales exponentially with the effective moduli-space dimensionality $D$, this corresponds to~$\langle {\cal T} \rangle$ also scaling 
exponentially in~$D$, which is compatible with the~\textsf{NP}-hard complexity class of the general problem~\cite{Denef:2006ad}. Optimal regions of the
landscape, however, have a mean residency time scaling at logarithmically in~$N_{\rm inf}$, and hence linearly in $D$. This does not contradict the~\textsf{NP}-hardness
classification --- \textsf{NP}-hardness is a worst-case assessment which does not preclude the existence of polynomial-time solutions for special instances
of the problem. Furthermore, the logarithmic divergence of the mean residency time signals a {\it dynamical phase transition}.  A similar non-equilibrium phase
transition occurs in quenched disordered media, when the probability distribution for waiting times reaches a critical power-law~\cite{disordered media}. 

\section{Landscape Dynamics as a Random Walk on a Network}
\label{RW complex nets}

The landscape can be modeled as a network (or graph) of nodes representing the different dS, AdS and Minkowski vacua. We assume that AdS and Minkowski vacua are terminal, acting as absorbing nodes. Network links, which define the network topology, represent the relevant transitions between vacua. For concreteness we assume these are governed by Coleman-De Luccia (CDL) instantons~\cite{Coleman:1977py,Callan:1977pt,Coleman:1980aw}.

As shown in the seminal papers of Garriga, Vilenkin and collaborators~\cite{Garriga:1997ef,Garriga:2005av}, a convenient approach to
study landscape dynamics is to follow a time-like geodesic (or ``watcher") in the eternally inflating space-time. See also~\cite{Nomura:2011dt,Garriga:2012bc}. 
In time, the watcher passes through a sequence of non-terminal vacua, until it finally hits a terminal vacuum. In the process, the watcher is performing a random
walk on the network of vacua. 

Let $N$ denote the total number of vacua in the network, taken to be comprised of $N_{\rm inf}$ inflating vacua and $N_{\rm term}$ terminal vacua.
In what follows, we will use capital indices $I,J = 1,\ldots,N$ for all vacua; indices $i,j = 1,\ldots, N_{\rm inf}$ for dS vacua, and $a,b = 1,\ldots,N_{\rm term}$ for
terminal (AdS and Minkowski) vacua. Unless otherwise stated, summations over
these indices are assumed to run over their respective range.

Let $f_i(\tau_i)$ denote the probability that the watcher is in vacuum $I$, as a function of the local proper time~$\tau_I$. 
Equivalently,~$f_I$ is the fraction of total comoving volume occupied by vacuum $I$. After coarse-graining over a
time interval~$\Delta \tau_I$ longer than transient evolution between periods of vacuum energy
domination, the change in occupation probability satisfies the master equation
\be
\Delta f_I = \sum_J  \left( \kappa^{\text{proper}}_{IJ} - \delta_{IJ} \sum_K \kappa^{\text{proper}}_{KJ} \right) \Delta\tau_J\, f_J \,,
\label{master0}
\ee
where $\kappa_{IJ}^{\text{proper}}$ is the $J \rightarrow I$ proper transition rate. The watcher's proper time is related to a global time
variable $t$ parametrizing the foliation through a lapse function~${\cal N}_I$~\cite{Garriga:2005av,Vanchurin:2006qp}:
\be
\Delta \tau_I = {\cal N}_I \Delta t \,.
\label{proper global} 
\ee
In a discrete setting, we think of~$t$ as a uniform discrete counter for transitions. In our analysis we will remain agnostic about the choice of time, as our goal is to define a time-reparametrization invariant measure.\footnote{A restriction on the choice of lapse function is that it should be well-defined in vacua of all types. For instance, scale-factor time, given by ${\cal N}_I = H_I^{-1}$, is
pathological in Minkowski vacua.} In terms of global time,~\eqref{master0} becomes
\be
\Delta f_I = \sum_J  \left( \kappa_{IJ} - \delta_{IJ} \sum_K \kappa_{KJ} \right) \Delta t\, f_J \,,
\label{master1}
\ee
where $\kappa_{IJ} \equiv \kappa^{\text{proper}}_{IJ} {\cal N}_J$. Therefore, in the continuum limit~\eqref{master1} becomes 
\be
\frac{{\rm d}f_I}{{\rm d}t} = \sum_{J} \mathbb{M}_{IJ}f_{J} \,,
\label{master}
\ee
where $\mathbb{M}_{IJ}$ is the transition matrix:
\be
\mathbb{M}_{IJ} \equiv \kappa_{IJ} - \delta_{IJ} \sum_K \kappa_{KJ} \,.
\label{M}
\ee
The sum over all rows of any column of $\mathbb{M}$ vanishes identically, $\sum_{I} \mathbb{M}_{IJ} = 0$, 
which enforces conservation of probability: $\sum_{I} f_I = 1$. The solution to~\eqref{master} is
\be
f(t) = {\rm e}^{\mathbb{M} t}f(0)\,,
\label{fsoln}
\ee
where $f(0)$ is the initial probability vector.

Since~$\mathbb{M}$ satisfies the sum rule~$\sum_{I} \mathbb{M}_{IJ} = 0$ and has positive off-diagonal elements, it 
follows from Perron-Frobenius' theorem that it has a single vanishing eigenvalue, while all other eigenvalues have strictly
negative real parts~\cite{Garriga:2005av}. (In fact we will see later that the non-zero eigenvalue with
largest real part is also non-degenerate and real.) The zero-eigensubspace is highly degenerate. Indeed, since the rate out of terminal
vacua vanishes by assumption, we have
\be
\kappa_{I a} = 0\,.
\label{kappa term 0}
\ee
Therefore, {\it any} vector lying in the terminal subspace, {\it i.e.}, of the form
\bea 
\nonumber
& & f_a^\infty \geq 0\qquad a \in {\rm terminals} \,;\\
& & f_i^\infty = 0\qquad i\in {\rm dS} \,,
\label{f stationary terms}
\eea
 is a zero-mode, $\mathbb{M} f^\infty = 0$. 
(This of course assumes there is at least one terminal vacuum. We will consider a toy landscape comprised only of dS vacua in Sec.~\ref{toy}.)
In particular, the solution~\eqref{fsoln} asymptotically tends to $f_a^\infty = f_a(0)$, $f_i^\infty = 0$. Thus the stationary distribution lies entirely
in the terminal subspace and is determined by initial conditions. 

\subsection{Occupational probability matrix}

It is convenient to express the probability vector in terms of the occupational probability matrix, or Green's function, $P_{IJ}(t)$.
This represents the probability that a random walker starting from $J$ at the initial time is at $I$ at time $t$. The probability vector can
be expanded as
\be
f_I(t) = \sum_{J} P_{IJ}(t) f_J(0)\,.
\label{f_i t}
\ee
Substituting into~\eqref{master}, the Green's function satisfies the master equation
\be
\frac{{\rm d}P_{IJ}}{{\rm d}t} = \sum_{K} \mathbb{M}_{IK}P_{KJ} \,;\qquad P_{IJ}(0) = \delta_{IJ}\,,
\label{forward_master}
\ee
with solution
\be
P_{IJ} (t) = \left({\rm e}^{\mathbb{M}t}\right)_{IJ}\,.
\label{P soln}
\ee

The master equation~\eqref{forward_master} can be decomposed into rate equations for terminal and non-terminal components. 
Setting $J = a$, it is easy to show that the solution is 
\be
P_{Ia} = \delta_{Ia}\,.
\label{PIalpha soln}
\ee
This is consistent with the general solution~\eqref{P soln}, together with $\sum_{K} \mathbb{M}_{IK}P_{Ka} =0$. Not surprisingly, a watcher
starting in a terminal vacuum must remain in that terminal forever. 

With~$J = j$, the rate equation breaks into
\begin{subequations} \label{P eqns}
\bea
\label{Pij eqn}
& \displaystyle \frac{{\rm d}P_{ij}}{{\rm d}t} = \sum_k M_{ik}P_{kj} \,; & \\
\label{Pai eqn}
& \displaystyle \frac{{\rm d}P_{a j}}{{\rm d}t} = \sum_i \kappa_{a i} P_{ij}  \,. &
\eea
\end{subequations}
The matrix $M$ appearing in~\eqref{Pij eqn} is an $N_{\rm inf} \times N_{\rm inf}$ square matrix defined by
\be
M_{ij} = \kappa_{ij} - \delta_{ij} \kappa_j \,,
\label{M def}
\ee
where $\kappa_j \equiv \sum_I \kappa_{Ij}$ is the total decay rate of vacuum~$j$.
The solution to~\eqref{Pij eqn} for $P_{ij}$ follows immediately:
\be
P_{ij} (t) = \left({\rm e}^{Mt}\right)_{ij}\,.
\label{PdS soln}
\ee

\subsection{CDL transition rates}
\label{transitions section}

To proceed we must be more specific about transition rates. For ${\rm dS}\rightarrow {\rm dS}$ transitions, the CDL rate is of the form
\be
\kappa_{ij} = \frac{A_{ij}}{w_j} \,.
\label{kappa dSdS}
\ee
Here $A_{ij} = \left(\Lambda^4 {\rm e}^{-S_{\rm bounce}}\right)_{ij}$ is the adjacency matrix, with $S_{\rm bounce}$ denoting the Euclidean action
of the bounce solution and $\Lambda^4$ the fluctuation determinant. The important property for our purposes is
that this matrix is symmetric~\cite{Lee:1987qc}:
\be
A_{ij} = A_{ji}\,.
\ee
The other factor in~\eqref{kappa dSdS} is the weight~$w_j$ of the parent dS vacuum:
\be
w_j = H_j^3 {\cal N}_j^{-1} {\rm e}^{S_j}\,,
\label{weights}
\ee
where $S_j = 48\pi^2M_{\rm Pl}^4/V_j$ is the dS entropy of the parent vacuum. The factor of $H_j^3 = \left(V_j/3M_{\rm Pl}^2\right)^{3/2}$
converts the CDL rate per unit volume to a transition rate, while the factor of~${\cal N}_j^{-1}$ converts the rate from unit proper time to global time
via~\eqref{proper global}. 

Transitions from inflating to terminal vacua are also assumed to be governed by CDL instantons. In this case the
${\rm dS}\rightarrow {\rm AdS/Minkowski}$ rate is given by
\be
\kappa_{a j} = \frac{\left(\Lambda^4 {\rm e}^{-S_{\rm bounce}}\right)_{a j}}{w_j}\,.
\label{kappa dSAdS}
\ee
This is similar to~\eqref{kappa dSdS}, except of course that the numerator is no longer symmetric.

\subsection{On time-reparametrization invariance}
\label{time reparam}

Since our goal is to define a time-reparametrization invariant measure, it is important to identify the invariant building blocks.
First note from~\eqref{weights} that the combination  
\be
\omega_j \equiv \frac{w_j}{\Delta t}  = \frac{H_j^3 {\rm e}^{S_j}}{\Delta\tau_j} 
\label{weight inv}
\ee
is gauge invariant. Similarly, the rates~\eqref{kappa dSdS} and~\eqref{kappa dSAdS} are both inversely proportional to~$w_j$,
hence the dimensionless transition probability
\be
\kappa_{Ij} \Delta t = \kappa_{Ij}^{\rm proper} \Delta\tau_j 
\ee
is also invariant. Thus the transition probability matrix~$M_{ij} \Delta t$ has time-reparametrization invariant elements: 
\be
M_{ij} \Delta t = \left(\kappa_{ij} - \delta_{ij} \kappa_j\right)\Delta t = \left(\kappa_{ij}^{\rm proper} - \delta_{ij} \kappa_j^{\rm proper}\right)  \Delta\tau_j \,,
\ee
where $\kappa_j \equiv \sum_I \kappa_{Ij}$ is the total decay rate of vacuum $j$. In particular the eigenvectors and
eigenvalues of~$M_{ij} \Delta t$ are invariant.

\subsection{Detailed balance and downward approximation}

A key feature of the dS-dS transition rate~\eqref{kappa dSdS} is that it is the ratio of a symmetric matrix and a
weight factor proportional to the exponential of the dS entropy. (Our results apply to any rate, CDL or
otherwise, of this form.) An immediate consequence is that the rates satisfy detailed balance,
\be
\frac{\kappa_{ji}}{\kappa_{ij}} = \frac{w_j}{w_i}\sim {\rm e}^{S_j-S_i}\,.
\label{detailed balance}
\ee
Thus upward tunneling is suppressed compared to downward tunneling by an exponential of the difference in dS entropy.
Importantly,~\eqref{detailed balance} only depends on the false and true vacuum potential energy. It is insensitive to details of the potential barrier
and does not rely on the thin-wall approximation.

This allows one to define a ``downward" approximation in which upward tunneling is neglected to leading order~\cite{SchwartzPerlov:2006hi,Olum:2007yk}.
By labeling dS vacua in order of increasing potential energy, $0 < V_1 \leq \ldots \leq V_{N_{\rm inf}}$, the transition matrix $M_{ij}$ for dS vacua
defined in~\eqref{M def} takes the form
\be
M = \begin{bmatrix}
-\kappa_1 & \kappa_{12} & \kappa_{13} &  \ldots  \\
0 & -\kappa_2 & \kappa_{23} &  \ldots \\
 \vdots & 0 & \ddots  &  \\
0 & 0  & 0 & -\kappa_{N_{\rm inf}} 
\end{bmatrix} + M_{\rm up}\,.
\label{M original}
\ee
(Recall that $\kappa_j \equiv \sum_I \kappa_{Ij}$ is the total decay rate of~$j$.) The leading, upper-triangular matrix encodes all downward transitions. The second term, $M_{\rm up}$, encodes all (exponentially small) upward transitions. In the ``downward" approximation~\cite{SchwartzPerlov:2006hi,Olum:2007yk}, one treats $M_{\rm up}$ perturbatively. We will make
use of this approximation later on to simplify some of our results. 

\subsection{Spectral analysis}

Although $M$ is not symmetric, it nevertheless has real eigenvalues, and its eigenvectors form a complete basis  of the $N_{\rm inf}$-dimensional subspace of dS vacua.
To see this, define the auxiliary matrix
\be
\Sigma = W^{-1/2} M \, W^{1/2}\,;\qquad  W \equiv {\rm diag}(\omega_1,\omega_2,\ldots,\omega_{N_{\rm inf}})\,,
\label{Sig M}
\ee
where~$\omega_i = w_i/\Delta t$ is the invariant weight defined in~\eqref{weight inv}. Using~\eqref{kappa dSdS} for the rate between dS vacua, it is straightforward
to see that~$\Sigma$ is symmetric and therefore has real eigenvalues. Moreover, since~\eqref{Sig M} defines a similarity transformation,~$\Sigma$
and~$M$ have identical spectra. Importantly, per the discussion in Sec.~\ref{time reparam}, the matrix~$\Sigma\Delta t \equiv W^{-1/2} M\Delta t \, W^{1/2} $
is manifestly time-reparametrization invariant, and therefore so are its eigenvalues and eigenvectors.  

For simplicity we assume that $M$ is irreducible, {\it i.e.}, there exists a sequence of transitions connecting any pair of inflating vacua. 
It then follows from Perron-Frobenius' theorem that its largest eigenvalue is non-degenerate and negative,
$\lambda_1 \leq 0$, while all other eigenvalues are strictly smaller. In other words,
\be
0\geq \lambda_1 > \lambda_2 \geq \ldots \geq \lambda_{N_{\rm inf}}\,.
\label{lamba's M}
\ee
Furthermore, $\lambda_1 = 0$ if and only if the decay rate into terminals vanishes for all dS vacua. Incidentally, we have already seen
that the full transition matrix~$\mathbb{M}$ has a single vanishing eigenvalue, with $N_{\rm term}$ degeneracy, while all other~$N_{\rm inf}$ 
eigenvalues have strictly negative real parts. It is easy to show, given the form of $\mathbb{M}$, that the latter set of eigenvalues coincide
with~\eqref{lamba's M}. Therefore~$\mathbb{M}$ has real eigenvalues, and its non-zero eigenvalues coincide with those of~$M$. In particular,
its largest, non-vanishing eigenvalue is~$\lambda_1$, which sets the relaxation time for the Markov process.

The eigenvectors of $\Sigma$, denoted by~$v^{(\ell)}$, $\ell = 1,\ldots, N_{\rm inf}$, form a complete, orthonormal and gauge invariant basis
of the $N_{\rm inf}$-dimensional subspace of dS vacua:
\be
\sum_{\ell =1}^{N_{\rm inf}} v^{(\ell)}_i v^{(\ell)}_j  = \delta_{ij} \,;\qquad \sum_i v^{(\ell)}_i v^{(\ell')}_i = \delta^{\ell \ell'}\,.
\label{comp ortho}
\ee
We mention in passing a further consequence of Perron-Frobenius' theorem, namely that the components of the dominant eigenvector can be chosen to all be positive:
\be
v^{(1)}_i \geq 0\,.
\label{v1 > 0}
\ee
Meanwhile, the eigenvectors of~$M$ are simply related to those of $\Sigma$ via 
\be
v^{(\ell)}_M = W^{1/2}v^{(\ell)}\,.
\label{vM v}
\ee
Thus the eigenvectors of $M$ also form a complete basis, albeit not orthonormal. Furthermore, they are simply related to those of~$\mathbb{M}$,
with eigenvalues given by~\eqref{lamba's M}. Indeed, it is easy to show, given the form of~$\mathbb{M}$, that the corresponding eigenvectors are:
\be
v^{(\ell)}_{\mathbb{M}\,i} = v^{(\ell)}_{M\,i} \,;\qquad v^{(\ell)}_{\mathbb{M}\,a} = \frac{1}{\lambda_\ell} \sum_i \kappa_{a i} v^{(\ell)}_{M\,i}\,.
\label{vbig M}
\ee

Given the form~\eqref{M def} of~$M_{ij}$, it is straightforward to show that its dominant eigenvalue is given by
\be
\lambda_1 = - \sum_{a = 1}^{N_{\rm term}} \frac{\sum_i \kappa_{ai} \sqrt{\omega_i}v_i^{(1)}}{\sum_i \sqrt{\omega_i}v_i^{(1)}} \,.
\label{lambda 1 explicit} 
\ee
In the downward approximation~\cite{SchwartzPerlov:2006hi}, in particular, one neglects upward transitions to leading order,
and the transition matrix~\eqref{M original} becomes upper-triangular. Hence its eigenvalues are simply given by its
diagonal elements. It follows that~$\lambda_1$ corresponds to the smallest decay rate~\cite{SchwartzPerlov:2006hi}:
\be
\lambda_1 \simeq -{\rm min}\left\{\kappa_j\right\}  \qquad (\text{downward})\,.
\label{downward lambda 1}
\ee
Thus the relaxation time is determined by the longest-lived dS vacuum. Similarly, in this approximation the other eigenvalues $\lambda_2,\ldots, \lambda_{N_{\rm inf}}$ are given by the decay rates of dS vacua in increasing order of instability.

In terms of the eigenvalues and eigenvectors of~$\Sigma$, the solution~\eqref{PdS soln} for $P_{ij}$ becomes
\be
P_{ij} (t)  = \left(W^{1/2} {\rm e}^{\Sigma t} W^{-1/2}\right)_{ij} = \sqrt{\frac{\omega_i}{\omega_j}}\, \sum_{\ell=1}^{N_{\rm inf}} {\rm e}^{\lambda_\ell t} \, v^{(\ell)}_i v^{(\ell)}_j \,.
\label{Pij soln}
\ee
By orthonormality~\eqref{comp ortho} this satisfies the correct initial condition, $P_{ij}(0) = \sqrt{\omega_i/\omega_j} \sum_\ell v^{(\ell)}_i v^{(\ell)}_j  = \delta_{ij}$.
When computing first-passage statistics in Sec.~\ref{MFPT section} we will make use of the Laplace
transform of~\eqref{Pij soln}, defined as usual by $\tilde{P}_{ij}(s) = \int_0^\infty {\rm d}t\, P_{ij}(t) {\rm e}^{-st}$. The result is
\be
\tilde{P}_{ij}(s) = \sqrt{\frac{\omega_i}{\omega_j}} \,\sum_{\ell=1}^{N_{\rm inf}} \frac{v^{(\ell)}_i v^{(\ell)}_j}{s -\lambda_\ell}\,.
\label{Pij soln Lap}
\ee

Next we can solve for $P_{a j}$ by substituting~\eqref{Pij soln} into the master equation~\eqref{Pai eqn}:
\be
\frac{{\rm d}P_{a j}}{{\rm d}t} =  \sum_{\ell} {\rm e}^{\lambda_\ell t} \sum_i \kappa_{a i} \sqrt{\frac{\omega_i}{\omega_j}}v^{(\ell)}_i v^{(\ell)}_j \,.
\label{dPaj}
\ee
The solution with initial condition $P_{a j}(0) = 0$ is
\be
P_{a j}(t) = \sum_{\ell} \frac{{\rm e}^{\lambda_\ell t}-1}{\lambda_\ell}  \sum_i \kappa_{a i} \sqrt{\frac{\omega_i}{\omega_j}}v^{(\ell)}_i v^{(\ell)}_j \,.
\label{Paj soln}
\ee
Equations~\eqref{PIalpha soln},~\eqref{Pij soln} and~\eqref{Paj soln} form the solution for the occupational probability matrix $P_{IJ}(t)$.

\subsection{dS-only toy landscape}
\label{toy}

Our measure will be simplest to define in the case of a toy landscape comprised of dS vacua only, {\it i.e.}, without terminals. This will be the focus of Sec.~\ref{measure dS}.
With this in mind, we collect here a few useful results for dS-only landscapes. 

In the absence of terminals, the transition matrix~$M_{ij}$ is still given by~\eqref{M def}, but with~$\kappa_j = \sum_k \kappa_{kj}$.
Hence in this case the sum of any column of~$M$ vanishes,~$\sum_i M_{ij} = 0$. It then follows from Perron-Frobenius' theorem that the largest eigenvalue
of~$M$ is non-degenerate and vanishes:
\be
\lambda_1 = 0\,. 
\ee
The Green's function~\eqref{Pij soln} therefore reduces to 
\be
P_{ij} (t)  = \sqrt{\frac{\omega_i}{\omega_j}}\left(v^{(1)}_i v^{(1)}_j  +  \sum_{\ell\geq 2} {\rm e}^{\lambda_\ell t} \, v^{(\ell)}_i v^{(\ell)}_j\right) \,,
\label{Pij soln dS only}
\ee
where we have isolated the zero-mode for convenience. Its Laplace transform is 
\be
\tilde{P}_{ij}(s) = \sqrt{\frac{\omega_i}{\omega_j}} \left(\frac{v^{(1)}_i v^{(1)}_j}{s} +  \sum_{\ell\geq 2} \frac{v^{(\ell)}_i v^{(\ell)}_j}{s - \lambda_{\ell}} \right)\,.
\label{Pij soln Lap dS only}
\ee

It is straightforward to derive an explicit expression for~$v^{(1)}$ and~$v^{(1)}_{M}$, the
zero-modes for $\Sigma$ and $M$ respectively. Note that $v^{(1)}_{M}$, in particular, sets the stationary
distribution: $\frac{{\rm d}f^\infty}{{\rm d}t} = M f^\infty = 0$. To do so, first write 
\be
M = ZW^{-1}\,;\qquad Z_{ij} \equiv A_{ij} -  \delta_{ij} \sum_r A_{rj}\,,
\label{M simple}
\ee
where $A_{ij}$ is the symmetric matrix defined in~\eqref{kappa dSdS}. Substituting into~\eqref{Sig M} gives
\be
\Sigma = W^{-1/2} M \, W^{1/2} = W^{-1/2}ZW^{-1/2}\,.
\label{Sig Z}
\ee
Now, notice that the vector with unit entries, $\vec{e} \equiv (1,1,\ldots,1)$, is a zero-eigenvector of $Z$. It follows that
$\Sigma \left(W^{1/2} \vec{e}\right) = 0$, hence $v^{(1)} \sim W^{1/2} \vec{e}$. Normalizing, we obtain
\be
v^{(1)}_i = \sqrt{\frac{\omega_i}{\omega}}\,;\qquad \omega \equiv \sum_i \omega_i\,.
\label{zero mode Sig}
\ee
Per~\eqref{vM v}, the corresponding zero-mode of $M$ is 
\be
v^{(1)}_{M\,i} = \sqrt{\omega_i} v^{(1)}_i = \frac{\omega_i}{\sqrt{\omega}} \,.
\ee
The stationary distribution $f^\infty$ is proportional to $v^{(1)}_{M}$, and by definition satisfies $\sum_i f^\infty_i = 1$.
It follows that
\be
f^\infty_i  =  \frac{w_i}{w} \sim v^{(1)\,2}_i\,.
\label{finf}
\ee
From this point of view, $v^{(1)}$ can be thought as a gauge invariant generalization of the (dS-only) stationary measure~$f^\infty$. 

\section{First-Passage Statistics}
\label{MFPT section}

The key building block in defining the accessibility measure is the mean first-passage time (MFPT). This is the average time
taken by a random walker starting from a given initial node to reach a given target for the first time. The MFPT
has been applied to random walks in various contexts~\cite{MFPT book} and is a standard measure
of search efficiency on networks, {\it e.g.},~\cite{MFPT ref}. Notably, in cosmology first-passage statistics have been used in the context of stochastic inflation~\cite{Vennin:2015hra,Assadullahi:2016gkk,Vennin:2016wnk}, in particular to study tunneling between vacua~\cite{Noorbala:2018zlv}. The MFPT was applied to landscape dynamics in~\cite{Khoury:2019yoo}, and some of the concepts covered below also appeared in~\cite{Khoury:2019yoo}.

\subsection{First-passage density}

A central quantity in first-passage statistics is the {\it first-passage density}, $F_{IJ}(t)$, $I\neq J$. This is the probability density
that a random walker who started at node $J$ at $t=0$ visits node $I$ for the first time at time $t$. All first-passage
statistics can be derived from $F$. For instance, the MFPT is given by its first moment:
\be
\langle t_{J\rightarrow I}\rangle = \frac{\int_0^\infty {\rm d}t\,t \,F_{IJ}(t)}{\int_0^\infty {\rm d}t \,F_{IJ}(t)} =  - \frac{{\rm d}\ln\tilde{F}_{IJ}(s)}{{\rm d}s}\bigg\vert_{s = 0} \,.
\label{MFPT def}
\ee
Thus $\langle t_{J\rightarrow I}\rangle$ is the average time for a random walker starting from $J$ to reach $I$,
averaged over all paths connecting the two nodes.

A well-known equation relating first-passage density and occupational probability is~\cite{MFPT book}
\be
P_{IJ}(t) = \int_0^t {\rm d}t'\, F_{IJ}(t') P_{II}(t-t')\,; \qquad I\neq J\,.
\label{famous}
\ee
The meaning of this relation is clear: the occupational probability at time $t$ is given by the probability that the walker
has reached $I$ at any earlier time $t'$ multiplied by the ``loop" probability $P_{II}$ that the walker returned to $I$
in the remaining time $t-t'$. Clearly $F_{IJ}$ is non-vanishing only if the initial node is a dS vacuum, thus we henceforth set $J = j$.
It remains to specify whether the final node $I$ is a terminal or non-terminal vacuum.

Consider first the case where the final node is a terminal vacuum ($I = a$). Using $P_{aa}(t-t') = 1$, which follows from~\eqref{PIalpha soln},
we can differentiate~\eqref{famous} to obtain
\be
F_{a j}(t) = \frac{{\rm d}P_{a j}}{{\rm d}t} =  \sum_{\ell} {\rm e}^{\lambda_\ell t} \sum_i \kappa_{a i} \sqrt{\frac{\omega_i}{\omega_j}}v^{(\ell)}_i v^{(\ell)}_j\,,
\label{first hit term 0}
\ee
where in the last step we have substituted~\eqref{dPaj}. The Laplace transform of this result, which will be useful later on, is
\be
\tilde{F}_{a j}(s) =  \sum_{\ell} \frac{1}{s-\lambda_\ell} \sum_i \kappa_{a i} \sqrt{\frac{\omega_i}{\omega_j}}v^{(\ell)}_i v^{(\ell)}_j = \sum_{\ell} \frac{\lambda_\ell}{s-\lambda_\ell} \frac{v^{(\ell)}_j}{\sqrt{\omega_j}}v^{(\ell)}_{\mathbb{M}\,a}\,,
\label{first hit term}
\ee
where we have used~\eqref{vbig M}. Differentiating and setting $s = 0$ gives the MFPT~\eqref{MFPT def} from dS vacuum $i$ to terminal $a$:
\be
\langle t_{j\rightarrow a}\rangle = - \frac{{\rm d}\ln\tilde{F}_{aj}(s)}{{\rm d}s}\bigg\vert_{s = 0} 
=  \frac{\sum_{\ell} \left\vert\lambda_\ell\right\vert^{-1} v^{(\ell)}_j v^{(\ell)}_{\mathbb{M}\,a}}   {\sum_{\ell'}v^{(\ell')}_j\,v^{(\ell')}_{\mathbb{M}\,a}}\,.
\ee

Consider next the situation where the final node is a dS vacuum ($I = i$). In this case,~\eqref{famous} becomes  
\be
P_{ij}(t) = \int_0^t {\rm d}t'\, F_{ij}(t') P_{ii}(t-t')\,; \qquad i\neq j\,.
\label{famous dS}
\ee
Taking the Laplace transform we obtain
\be
\tilde{F}_{ij}(s) = \frac{\tilde{P}_{ij}(s)}{\tilde{P}_{ii}(s)} \,.
\label{first hit dS}
\ee
Substituting~\eqref{Pij soln Lap} gives the desired result
\be
\tilde{F}_{ij}(s)  =  \sqrt{\frac{\omega_i}{\omega_j}}\, \frac{v^{(1)}_i v^{(1)}_j + (s-\lambda_1)\sum_{\ell\geq 2}\frac{1}{s -\lambda_\ell} v^{(\ell)}_i v^{(\ell)}_j}{v^{(1)\,2}_i + (s-\lambda_1)\sum_{\ell'\geq 2} \frac{1}{s - \lambda_{\ell'}} v^{(\ell')\,2}_i}\,.
\label{first hit dS explicit}
\ee

In particular, let us focus on the special case of a toy landscape without terminals studied in Sec.~\ref{toy}. Setting~$\lambda_1 = 0$ and using~\eqref{zero mode Sig} gives 
\be
\tilde{F}_{ij}(s)  = \frac{v^{(1)\,2}_i + s \frac{v^{(1)}_i}{v^{(1)}_j} \sum_{\ell\geq 2}\frac{1}{s -\lambda_\ell} v^{(\ell)}_i v^{(\ell)}_j}{v^{(1)\,2}_i + s\sum_{\ell'\geq 2} \frac{1}{s - \lambda_{\ell'}} v^{(\ell')\,2}_i}\qquad (\text{dS only})\,.
\label{first hit dS only}
\ee
Note that for finite $N_{\rm inf}$ the ever-hitting probability, $\tilde{F}_{ij}(0) =  \int_0^\infty {\rm d}t F_{ij}(t)$, is unity in this case. (This is in contrast with~\eqref{first hit dS explicit}, whose
ever-hitting probability is less than unity due to the loss of probability to terminals.) Differentiating and setting $s = 0$ gives the MFPT between two dS vacua in a landscape without terminals:
\be
\langle t_{j \rightarrow i}\rangle = \frac{1}{v^{(1)\,2}_i} \sum_{\ell= 2}^{N_{\rm inf}} \frac{1}{|\lambda_\ell|} \left( v^{(\ell)\,2}_i - \frac{v^{(1)}_i}{v^{(1)}_j}  v^{(\ell)}_i v^{(\ell)}_j\right) \qquad (\text{dS only})\,.
\label{dSdS MFPT}
\ee
We will make use of this result when defining the accessibility measure in Sec.~\ref{measure dS}.

\subsection{First-return density}
\label{1st return}

One can similarly define a {\it first-return density}, $F_{II}(t)$, which represents the probability density that a random walker returns
at the initial node at time~$t$. Clearly this is only non-trivial if the node in question is a dS vacuum, hence we set $I = i$. 

We proceed by generalizing~\eqref{famous dS} to allow for the initial and final nodes to be the same. To do so, it is convenient
to discretize time in units of $\Delta t$, where $\Delta t$ is the coarse-graining time interval~\eqref{proper global}. That is,~$t = n \Delta t$, with~$n$ an integer.
The generalization of~\eqref{famous} is then~\cite{Khoury:2019yoo}
\be
P_{ij}(n) = \delta_{ij}\delta_{n0} +\sum_{m=0}^n \,F_{ij}(m) P_{ii}(n-m) \Delta t\,,
\label{famous 2}
\ee
where we have dropped corrections of $\mathcal{O}\left((\Delta t)^2\right)$. Taking the discrete Laplace transform, defined as $\tilde{f}(s) = \sum_{n=0}^\infty f(n) {\rm e}^{-s n\Delta t} \Delta t$, gives 
\be
\tilde{P}_{ij}(s) = \delta_{ij}\Delta t + \tilde{F}_{ij}(s)\tilde{P}_{ii}(s)\,,
\label{PFs}
\ee
which agrees with~\eqref{first hit dS} for $i \neq j$. For the case of interest, $i=j$, we obtain
\be
\tilde{F}_{ii}(s) = 1 - \frac{\Delta t}{\tilde{P}_{ii}(s)}\,.
\label{first return dS}
\ee
Substituting~\eqref{Pij soln Lap} gives
\be
\tilde{F}_{ii}(s) = 1 - \frac{(s-\lambda_1) \,\Delta t}{v^{(1)\,2}_i + (s-\lambda_1) \sum_{\ell\geq 2} \frac{1}{s - \lambda_{\ell}} v^{(\ell)\,2}_i} \,.
\label{first return dS 2}
\ee

In the special case of a toy landscape without terminal vacua, we can set $\lambda_1 = 0$ to obtain
\be
\tilde{F}_{ii}(s) = 1 - \frac{s\,\Delta t}{v^{(1)\,2}_i + s \sum_{\ell\geq 2} \frac{1}{s - \lambda_{\ell}} v^{(\ell)\,2}_i} \qquad (\text{dS only})\,.
\label{first return dS only}
\ee
For finite $N_{\rm inf}$, this implies that the ever-return probability, $\tilde{F}_{ii}(0) =  \int_0^\infty {\rm d}t F_{ii}(t)$, is unity. 
Meanwhile, the first moment of this distribution gives the mean first-return time: 
\be
\langle t_{i \rightarrow i}\rangle = - \frac{{\rm d}\tilde{F}_{ii} (s)}{{\rm d}s}\bigg\vert_{s = 0} = \frac{\Delta t}{v^{(1)\,2}_i} \qquad (\text{dS only})\,.
\label{kac lemma}
\ee
Thus the mean first-return time is set by the stationary distribution, which is Kac's celebrated lemma~\cite{Kac}.

\section{Accessibility Measure Without Terminals}
\label{measure dS}

We are now in a position to define the accessibility measure. The measure is conceptually easiest to define in the absence of terminal vacua, {\it i.e.}, in a toy 
landscape comprised of dS vacua only. This is the subject of this Section. In Sec.~\ref{measure with terms} we will generalize the measure to include
terminals. 

The key building block is a weighted MFPT, defined in Sec.~\ref{partial MFPT}, called the partial MFPT (pMFPT). The pMFPT can be expressed simply in terms of the eigenvalues and eigenvectors of the transition matrix. Moreover, it admits other equivalent representations, each offering different insights. Firstly, we will show in Sec.~\ref{pMFPT 1st return} that the pMFPT can be neatly expressed in terms of first-return statistics. This form makes manifest that the pMFPT is independent of initial conditions, and immediately implies an important lower bound for the pMFPT.  Secondly we will show in Sec.~\ref{pMFPT no miss} that minimizing the pMFPT is equivalent to maximizing the first-passage probability at intermediate times. Thirdly, we will see in Sec.~\ref{escape} can be related to the escape or never-return probability. The accessibility measure is then defined in Sec.~\ref{measure dS def} in terms of the reciprocal of the pMFPT.

\subsection{pMFPT}
\label{partial MFPT}

We define the dimensionless {\it partial MFPT (pMFPT)} to the dS vacuum $j$ by
\be
{\cal T}_i \equiv \frac{v^{(1)\,2}_i}{1+v^{(1)\,2}_i} \sum_{j \neq i} v^{(1)\,2}_j  \frac{\langle t_{j\rightarrow i}\rangle}{\Delta t} \,,
\label{pMFPT def}
\ee 
where $\langle t_{j\rightarrow i}\rangle$ is the MFPT~\eqref{dSdS MFPT} between dS vacua $j$ and $i$,
and $\Delta t$ is the coarse-graining time interval.\footnote{The partial MFPT is closely related to a quantity
first defined in~\cite{GMFPT}, which the authors called the global MFPT. We prefer the term partial MFPT,
since, as we will see~\eqref{kemeny 1}, summing~${\cal T}_i$ over $i$ gives Kemeny's constant,~${\cal T}_{\rm MFPT}$,
which to our mind is a more apt measure of global MFPT.} Since~$v^{(1)\,2}_j$ sets the stationary distribution, per~\eqref{finf},
~${\cal T}_i$ can be thought of as the MFPT to node~$i$, averaged over all initial nodes weighted by the stationary distribution. 
The overall factor of~$\frac{v^{(1)\,2}_i}{1+v^{(1)\,2}_i}$ is included to simplify some of the resulting expressions.

As mentioned in the Introduction, weighing the MFPT's $\langle t_{j\rightarrow i}\rangle$ by the stationary distribution yields a conservative
estimate of the characteristic time needed to reach~$i$. Indeed, the stationary distribution~$v^{(1)\,2}_j$ exponentially
favors the lowest-lying dS vacuum, as can be seen from~\eqref{zero mode Sig}, which in turn is the most stable vacuum. (To leading order
in the downward approximation, this vacuum is absolutely stable.) The corresponding MFPT~$\langle t_{j\rightarrow i}\rangle$ should be the largest.
Thus weighing with the stationary distribution amounts to giving greatest weight to the initial node with the longest average travel time to~$i$.
It also stands to reason that any other distribution with non-zero support on the most stable dS vacuum, such as the uniform distribution, 
should yield similar results. We leave to future work a detailed study of sensitivity to different distributions and proceed for now with~\eqref{pMFPT def}
as a definition of the pMFPT. 

Substituting~\eqref{dSdS MFPT} for $\langle t_{j\rightarrow i}\rangle$, and using the identity $\sum_{j\neq i} v_j^{(1)}v_j^{(\ell)} = -  v_i^{(1)}v_i^{(\ell)}$ (which follows from the orthonormality relation~\eqref{comp ortho}), we obtain
\be
{\cal T}_i  = \sum_{\ell= 2}^{N_{\rm inf}} \frac{v^{(\ell)\,2}_i}{|\lambda_\ell| \Delta t}  \,.
\label{pMFPT 2}
\ee
Thus the pMFPT is simply related to the non-zero eigenvalues and corresponding eigenvectors (`relaxing modes') of the transition matrix.
The eigenvectors are time-reparametrization invariant, as mentioned earlier. Meanwhile, the $\lambda_\ell \Delta t$'s are eigenvalues of the 
gauge invariant probability matrix $\Sigma\Delta t = W^{-1/2} M\Delta t \, W^{1/2}$. {\it Therefore the pFMPT is manifestly
time-reparametrization invariant.} 

Importantly, the eigenvalues and eigenvectors are to an excellent approximation determined by the leading upper-triangular matrix in~\eqref{M original},
encoding downward transitions, up to exponentially small corrections due to~$M_{\rm up}$. As such, unlike stationary measures, the pMFPT is robust
against small tweaks to the landscape, of the kind discussed in the Appendix. Finally, we note in passing that summing~\eqref{pMFPT 2} over~$i$ gives a global MFPT, otherwise known as Kemeny's constant~\cite{kemeny}:
\be
{\cal T}_{\rm MFPT} = \sum_i {\cal T}_i  = \sum_{\ell= 2}^{N_{\rm inf}} \frac{1}{|\lambda_\ell|\Delta t}\,.
\label{kemeny 1}
\ee
This matches the MFPT studied in~\cite{Khoury:2019yoo} for finite regions of the landscape. 

\subsection{pMFPT and first-return statistics}
\label{pMFPT 1st return}

The pMFPT can be neatly expressed in terms of first-return statistics~\cite{GMFPT}. Consider the second moment of the first-return probability,
\be
\langle t_{i\rightarrow i}^{\,2}\rangle = \int_0^\infty {\rm d}t\,t^2 F_{ii}(t) = \frac{{\rm d}^2\tilde{F}_{ii}(s)}{{\rm d}s^2}\bigg\vert_{s = 0}\,,
\ee
where we have used the fact that the first-return probability for dS-only vacua, given by~\eqref{first return dS only}, is normalized: $\int_0^\infty {\rm d}t\,F_{ii}(t) = \tilde{F}_{ii}(0) = 1$.
Substituting~\eqref{first return dS only}, we obtain
\be
\langle t_{i\rightarrow i}^{\,2}\rangle = 2 \frac{\Delta t}{v^{(1)\,4}_i} \sum_{\ell= 2}^{N_{\rm inf}} \frac{v^{(\ell)\,2}_i}{|\lambda_\ell|} = 2\langle t_{i\rightarrow i}\rangle^2  \sum_{\ell= 2}^{N_{\rm inf}} \frac{v^{(\ell)\,2}_i}{|\lambda_\ell|\Delta t}\,,
\ee
where the last step follows from Kac's lemma~\eqref{kac lemma}. Combining with~\eqref{pMFPT 2} gives the desired result:
\be
{\cal T}_i  = \frac{1}{2} \frac{\langle t_{i\rightarrow i}^{\,2}\rangle}{\langle t_{i\rightarrow i}\rangle^2}\,.
\label{pMFPT 3}
\ee
Thus the pMFPT is simply related to the variance of first-return times. Since first-return statistics by definition consider random walks that start at~$i$,
this expression makes clear that~${\cal T}_i$ is independent of initial conditions. Furthermore, since $\langle t_{i\rightarrow i}^{\,2}\rangle \geq \langle t_{i\rightarrow i}\rangle^2$,
this implies a lower bound for the pMFPT:
\be
{\cal T}_i  \geq \frac{1}{2}\,.
\label{pMFPT bound}
\ee
The inequality is saturated if the variance of first-return times vanishes, such that $\langle t_{i\rightarrow i}^{\,2}\rangle = \langle t_{i\rightarrow i}\rangle^2$.
This bound will play a key role in deriving phenomenological implications of the accessibility measure in Sec.~\ref{pheno}.

\subsection{pMFPT and first-passage probability}
\label{pMFPT no miss}

Perhaps the most intuitive interpretation of the pMFPT is through its relation to the first-passage probability. As we now show, minimizing ${\cal T}_i$ is equivalent
to maximizing the first-passage probability to~$i$ at early times compared to the relaxation time. 

Let us define the first-passage probability to~$i$ as a weighted average over initial nodes:
\be
H_i(t) = \sum_j v^{(1)\,2}_j \int_0^t {\rm d}t'\, F_{ij}(t')\,.
\ee
On a finite landscape, every vacuum is guaranteed to be accessed eventually, as reflected by the fact that~$H_i$ tends to unity as~$t\rightarrow \infty$:
\be
\lim_{t\rightarrow\infty} H_i(t) = \lim_{s \rightarrow 0} \sum_{j\neq i} v^{(1)\,2}_j\tilde{F}_{ij}(s) = \lim_{s \rightarrow 0} \frac{v^{(1)\,2}_i\left(1-s\sum_{\ell\geq 2} \frac{1}{s - \lambda_{\ell}} v^{(\ell)\,2}_i\right) }{v^{(1)\,2}_i + s\sum_{\ell'\geq 2} \frac{1}{s - \lambda_{\ell'}} v^{(\ell')\,2}_i} = 1\,,
\ee  
where we have substituted~\eqref{first hit dS only} and used orthonormality~\eqref{comp ortho}. Correspondingly, the probability that~$i$ has not yet been visited, $1 - H_i(t)$, starts from 1 and tends to 0 as~$t\rightarrow \infty$. While each vacuum is guaranteed to be populated eventually, the required time scale for {\it all} vacua to be accessed is of course the relaxation time. 

Our interest lies instead on the approach to equilibrium, {\it i.e.}, for~$t$ much smaller than the relaxation time.
To quantify the finite-time probability that a site~$i$ has not yet been visited, consider the simple figure of merit: 
\be
X_i \equiv \frac{1}{\Delta t} \int_0^\infty {\rm d}t \Big(1 - H_i(t)\Big)\,.
\ee
Nodes with small~$X_i$ have a higher probability of being accessed early, whereas those with high~$X_i$ tend to be populated later. 
In terms of Laplace transforms, we have
\be
X_i = \frac{1}{\Delta t} \lim_{s \rightarrow 0} \left(\frac{1}{s} - \tilde{H}_i(s)\right) =  \lim_{s \rightarrow 0}\left\{\frac{1}{s} \left( 1 -  \sum_j v^{(1)\,2}_j\tilde{F}_{ij}(s)\right)\right\} = \frac{1+v^{(1)\,2}_i}{v^{(1)\,2}_i} \sum_{\ell= 2}^{N_{\rm inf}} \frac{v^{(\ell)\,2}_i}{|\lambda_\ell| \Delta t}\,,
\ee
which implies the simple result
\be
X_i = \frac{1+v^{(1)\,2}_i}{v^{(1)\,2}_i}{\cal T}_i \,.
\ee
Hence, for fixed stationary distribution, nodes with smaller pMFPT have smaller $X_i$ and, correspondingly, higher probability of being accessed early on
relative to the relaxation time; in contrast, nodes with larger pMFPT are less likely to be accessed early on. 

\subsection{pMFPT and escape probability}
\label{escape}

Yet another useful expression for the pMFPT is in terms of the escape probability in the limit of an infinite landscape ($N_{\rm inf} \rightarrow \infty$).
First let us define the pseudo-Green's function~\cite{pseudogreen}:
\be
{\cal R}_i \equiv \frac{1}{\Delta t}\int_0^\infty {\rm d}t\,\Big( P_{ii}(t) - v^{(1)\,2}_i \Big) =  \frac{1}{\Delta t} \lim_{s\rightarrow 0} \left(\tilde{P}_{ii}(s) - \frac{v^{(1)\,2}_i}{s}\right)\,.
\label{pseudoGreen dS only}
\ee
Setting $i=j$ in~\eqref{Pij soln Lap dS only}, we have
\be
\tilde{P}_{ii}(s) = \frac{v^{(1)\,2}_i}{s} + \sum_{\ell\geq 2} \frac{v^{(\ell)\,2}_i}{s - \lambda_{\ell}} \,,
\label{Pii(s) dS only}
\ee
and it follows that
\be
{\cal R}_i = \sum_{\ell= 2}^{N_{\rm inf}} \frac{v^{(\ell)\,2}_i}{|\lambda_\ell| \Delta t} = {\cal T}_i\,.
\label{pseudoGreen dS only 2}
\ee
Thus ${\cal R}_i$ agrees with the pMFPT. 

On the other hand, the pseudo-Green's function can be related to the {\it escape probability} ${\cal S}_i$, defined as the probability that the
walker never returns to~$i$:
\be
{\cal S}_i \equiv 1 - \int_0^\infty {\rm d}t \,F_{ii}(t) = 1 - \lim_{s\rightarrow 0} \tilde{F}_{ii}(s)\,,
\label{escape prob}
\ee
where $\lim_{s\rightarrow 0} \tilde{F}_{ii}(s) = \int_0^\infty {\rm d}t\,F_{ii}(t)$ is the ever-return probability. The escape probability delineates whether random walks are {\it recurrent} at~$i$ (vanishing escape probability) or {\it transient} at~$i$ (finite escape probability): 
\bea
\nonumber
{\cal S}_i &=& 0 \qquad\;\;\;\;\;\; \Longleftrightarrow \qquad \text{recurrence at}~i \\ 
{\cal S}_i & = & {\rm finite} \qquad \Longleftrightarrow \qquad \text{transience at}~i \,.
\label{recur relation S}
\eea
Recurrence at~$i$ means that a random walker is certain to return eventually to~$i$, and, because the process is Markovian, will do so infinitely-many times in the future.

Substituting the first-return density~\eqref{first return dS}, we have
\be
{\cal S}_i  = \lim_{s\rightarrow 0} \frac{\Delta t}{\tilde{P}_{ii}(s)} \,,
\label{Si}
\ee
with $\tilde{P}_{ii}(s)$ given by~\eqref{Pii(s) dS only}. Therefore, whether random walks are recurrent or transient at $i$ depends on
whether $\lim_{s\rightarrow 0} \tilde{P}_{ii}(s)$ is divergent or finite, respectively. For finite~$N_{\rm inf}$, the first term in~\eqref{Pii(s) dS only}
diverges as~$s\rightarrow 0$, resulting in a divergent $\tilde{P}_{ii}$ and hence a vanishing escape probability. Therefore, not surprisingly, random walks
on finite networks are always recurrent. In the limit of an infinite network ($N_{\rm inf}\rightarrow \infty$), on the other hand, the first term in~\eqref{Pii(s) dS only} 
gives a vanishingly small contribution~\cite{Michelitsch:2017}, leaving us with
\be
\lim_{\substack{N_{\rm inf} \rightarrow \infty \\ s\rightarrow 0}} \frac{\tilde{P}_{ii}(s)}{\Delta t}  = \sum_{\ell\geq 2} \frac{v^{(\ell)\,2}_i}{|\lambda_{\ell}| \Delta t} \,.
\label{Pii s}
\ee
The order of limits matters --- one must first send $N_{\rm inf} \rightarrow \infty$ {\it before} taking the late-time ($s\rightarrow 0$) limit.
Equation~\eqref{Pii s} agrees with the pseudo-Green's function~${\cal R}_i$ (and therefore the pMFPT) defined at finite $N_{\rm inf}$. 

Combining~\eqref{Si} and~\eqref{Pii s} yields the desired relation between pMFPT and escape probability:
\be
{\cal S}_i = {\cal R}_i^{-1} = {\cal T}_i^{-1}\,,
\label{SRT}
\ee
where the large network limit is understood in calculating ${\cal S}_i$. Importantly, since the escape probability is
time-reparametrization invariant, as mentioned earlier,~\eqref{SRT} confirms that so is~${\cal T}_i$.  

\subsection{Accessibility measure}
\label{measure dS def} 

The accessibility measure is defined as the reciprocal of the pMFPT, suitably normalized:
\be
p_i \equiv \frac{{\cal T}_i^{-1}}{\sum_k {\cal T}_k^{-1}} = \frac{{\cal S}_i}{\sum_k {\cal S}_k}\,.
\label{pj dS def}
\ee
Since the pMFPT is both time-reparametrization invariant and independent of initial conditions, as argued above, 
so is~$p_i$. Furthermore, because the accessibility measure is constructed from first-passage statistics, it is clearly oblivious
to any comoving {\it vs} physical volume ambiguity. However, we cannot claim that the measure is unique, since the pMFPT
involved a choice of distribution to average over initial nodes, as discussed below~\eqref{pMFPT def}. 

The measure favors vacua that are easily accessed under time evolution, {\it i.e.}, vacua that
saturate~\eqref{pMFPT bound}: ${\cal T}_i \sim {\cal O}(1)$. As such, it is analogous to {\it closeness}
centrality~\cite{closeness1,closeness2}, a widely-used centrality index in studies of complex networks. 
The closeness measure assigns greater weight to nodes that can be reached on average with the
fewest number of steps. 

Unlike standard measures based on the stationary distribution, which exponentially favors a single
dominant vacuum, the accessibility measure allows for multiple vacua having comparable weight. Specificaly,
there can be many vacua that nearly saturate~\eqref{pMFPT bound}, {\it i.e.}, with ${\cal T}_i \sim {\cal O}(1)$,
and all will be weighted equally according to~\eqref{pj dS def}. Relatedly, unlike stationary measures, the accessibility
measure is insensitive to small tweaks to the landscape. It can be reliably calculated to leading order in the downward
approximation, where the transition matrix assumes the upper-triangular form~\eqref{M original}.

\section{Accessibility Measure With Terminals}
\label{measure with terms}

In the case of a dS-only landscape studied in the previous Section, we saw that the accessibility measure
could be expressed in four equivalent ways: 1.~In terms of a pMFPT, defined in~\eqref{pMFPT def} 
as a weighted average of the MFPT to a given node; 2.~In terms of the variance of first-return times
to the starting node; 3. In terms of a first-passage probability to a given node for times shorter than the relaxation time;  
4.~In terms of the escape probability from a given node. 

Once terminals are included, however, these give inequivalent definitions of the pMFPT. It turns out that the last approach in
terms of escape probability is most straightforward to generalize in a landscape with terminal
vacua. Specifically, in Sec.~\ref{pMFPT dS with terms} we will define the accessibility measure for dS vacua in terms of the
escape probability. In Sec.~\ref{pMFPT terms} we will instead define the accessibility measure for terminal vacua in terms of
a trapping probability. Both escape and trapping probabilities admit simple expressions in terms of eigenvectors and eigenvalues
of the transition matrix. Akin to the dS-only case, the resulting measure is both time-reparametrization invariant and
independent of initial conditions. 

\subsection{pMFPT and escape probability for dS vacua} 
\label{pMFPT dS with terms}

Following the steps in Sec.~\ref{escape}, we begin by defining the pseudo-Green's function~${\cal R}_i$. Analogously to~\eqref{pseudoGreen dS only},
the pseudo-Green's function is obtained by subtracting the dominant eigenvector contribution to the occupational probability:
\be
{\cal R}_i \equiv \frac{1}{\Delta t}\int_0^\infty {\rm d}t\,\Big( P_{ii}(t) - {\rm e}^{\lambda_1t}v^{(\ell)\,2}_i \Big) =  \frac{1}{\Delta t} \lim_{s\rightarrow 0} \left(\tilde{P}_{ii}(s) - \frac{v^{(1)\,2}_i}{s-\lambda_1}\right)\,.
\ee
Using~\eqref{Pij soln Lap} with $i=j$,
\be
\tilde{P}_{ii}(s) = \frac{v^{(1)\,2}_i}{s-\lambda_1}  + \sum_{\ell=2}^{N_{\rm inf}} \frac{v^{(\ell)\,2}_i}{s -\lambda_\ell} \,,
\label{Pii(s)} 
\ee
we obtain
\be
{\cal R}_i  =  \sum_{\ell=2}^{N_{\rm inf}} \frac{v^{(\ell)\,2}_i}{|\lambda_\ell|\Delta t}\,.
\label{pseudoGreen}
\ee
The result is identical in form to the dS-only pseudo-Green's function~\eqref{pseudoGreen dS only 2}, though of course the
eigenvalues and eigenvectors of the transition matrix are modified by the presence of terminals. In light of this we are led to {\it define}
the pMFPT for dS vacua as
\be
{\cal T}_i \equiv \sum_{\ell=2}^{N_{\rm inf}} \frac{v^{(\ell)\,2}_i}{|\lambda_\ell|\Delta t}\,.
\label{pMFPT dS def with terms}
\ee
In particular, summing over $i$ gives a measure of the global MFPT or Kemeny's constant analogous to~\eqref{kemeny 1}:
\be
{\cal T}_{\rm MFPT} = \sum_{i = 1}^{N_{\rm inf}} {\cal T}_i  = \sum_{\ell= 2}^{N_{\rm inf}} \frac{1}{|\lambda_\ell|\Delta t}\,.
\label{kemeny with terms}
\ee

The pseudo-Green's function is once again related to the escape probability~\eqref{escape prob},
\be
{\cal S}_i = 1 - \tilde{F}_{ii}(0) =  \frac{\Delta t}{\tilde{P}_{ii}(0)} \,.
\label{escape prob with terms}
\ee
As before, whether random walks are recurrent or transient at $i$ depends on whether $\lim_{s\rightarrow 0} \tilde{P}_{ii}(s)$ is
divergent or finite, respectively. Unlike the dS-only case, where the escape probability vanishes for finite $N_{\rm inf}$, in this case
the escape probability is finite due to the presence of terminals. In any case, in the limit of an infinite landscape ($N_{\rm inf}\rightarrow \infty$),
the first term in~\eqref{Pii(s)} gives a vanishingly small contribution~\cite{Michelitsch:2017}, resulting in
\be
{\cal S}_i = {\cal R}_i^{-1} =  {\cal T}_i^{-1}\,.
\label{SRT with terms}
\ee
Once again, the pseudo-Green's function coincides with the escape probability in the large-network limit.

Unlike the dS-only case, we have been unable to establish a strict lower bound on~${\cal T}_i$ akin
to~\eqref{pMFPT bound}. It is straightforward, however, to show that\footnote{To prove~\eqref{pMFPT dS almost lower bound},
define the renormalized Green's function, $P^{\rm R}_{ij}(t) \equiv {\rm e}^{-\lambda_1 t} P_{ij}(t)  = \sqrt{\omega_i/\omega_j} \sum_{\ell} {\rm e}^{(\lambda_\ell -\lambda_1) t} \, v^{(\ell)}_i v^{(\ell)}_j$, which remains finite asymptotically: $P^{\rm R}_{ij}(t \rightarrow\infty) = \sqrt{\omega_i/\omega_j}v^{(1)}_i v^{(1)}_j$. Thus its behavior is identical to the dS-only result~\eqref{toy}, except for the shift: $\lambda_\ell \rightarrow \lambda_\ell -\lambda_1$. Similarly, it follows from~\eqref{famous dS} that $P_{ij}^{\rm R}(t) = \int_0^t {\rm d}t'\, F_{ij}^{\rm R}(t') P_{ii}^{\rm R}(t-t')$, where
$F_{ij}^{\rm R}(t) = {\rm e}^{-\lambda_1 t}F_{ij}(t)$ is a renormalized first-passage density. In particular, the moments of the renormalized first-return density $F_{ii}^{\rm R}$ satisfy $\frac{1}{2} \frac{\left\langle t_{i\rightarrow i}^{{\rm R}\,2}\right\rangle}{\left\langle t_{i\rightarrow i}^{\rm R}\right\rangle^2} =\sum_{\ell= 2}^{N_{\rm inf}} \frac{v^{(\ell)\,2}_i}{(\lambda_1-\lambda_\ell)\Delta t}$, which implies~\eqref{pMFPT dS almost lower bound}.}
\be
\sum_{\ell= 2}^{N_{\rm inf}} \frac{v^{(\ell)\,2}_i}{(\lambda_1-\lambda_\ell)\Delta t} \geq \frac{1}{2}\,. 
\label{pMFPT dS almost lower bound}
\ee
In the limit $|\lambda_1| \ll |\lambda_\ell|$, corresponding to slow decay into terminals relative to dS-dS transitions,
this implies~${\cal T}_i \;\gsim\; 1/2$. For general~$\lambda_1$, we will show in Sec.~\ref{funnel sec} using the downward approximation
that the average pMFPT satisfies $\langle {\cal T}\rangle \;\gsim\; 1$ --- see~\eqref{avg T bound}. We have been
unable to generalize this to a strict lower bound for individual vacua.

\subsection{pMFPT and trapping probability for terminal vacua} 
\label{pMFPT terms}

For terminal vacua, the natural analogue of the escape probability is the trapping probability. Let us define a weighted trapping probability~${\cal S}_a$ 
as the late-time occupational probability at terminal node~$a$ averaged over all initial dS nodes~$j$:
\be
{\cal S}_a = |\lambda_1|\Delta t \lim_{t\rightarrow \infty} \frac{\sum_j P_{aj}(t) \sqrt{\omega_j}v_j^{(1)}}{\sum_i \sqrt{\omega_i}v_i^{(1)}}  \,.
\label{trap def}
\ee
This definition deserves some comments. The weighing factor~$\sqrt{\omega_i} v^{(1)}_i$ is
recognized as the dominant eigenvector~$v^{(1)}_{M\,i}$ of the transition matrix. Since $v^{(1)}_i \geq 0$, per~\eqref{v1 > 0},
this weighing factor is well-defined and, as we will see, leads to a simple expression for the trapping probability.
The overall factor of $|\lambda_1|\Delta t$ is included for convenience.

Recall from~\eqref{first hit term 0} the relation~$F_{a j}(t) = {\rm d}P_{a j}/{\rm d}t$ between first-passage and occupational probabilities.
Integrating this equation implies
\be
\lim_{t\rightarrow \infty} P_{aj}(t) = \tilde{F}_{aj}(0) = \sum_{\ell} \frac{1}{|\lambda_\ell|} \sum_i \kappa_{a i} \sqrt{\frac{\omega_i}{\omega_j}}v^{(\ell)}_i v^{(\ell)}_j \,,
\ee
where in the last step we have used~\eqref{first hit term}. Substituting this into~\eqref{trap def} and using the orthonormality of the eigenvectors,
we obtain
\be
{\cal S}_a = \frac{\sum_i \kappa_{ai}\Delta t \sqrt{\omega_i}v_i^{(1)}}{\sum_i \sqrt{\omega_i}v_i^{(1)}} \,.
\label{trap def 2}
\ee
This expression makes clear that the weighted trapping probability has a number of desirable properties: 

\begin{enumerate}

\item ${\cal S}_a$ is manifestly time-reparametrizaton invariant. 

\item Since the trapping probability is defined as a prescribed average over initial nodes, it is clearly independent of initial conditions.

\item As it should, ${\cal S}_a\rightarrow 0$ in the limit $\kappa_{ai}\rightarrow 0$, {\it i.e.}, if all transition rates to $a$ vanish. 

\item For node~$i$ to even qualify as a vacuum its decay rate should be less than an inverse unit time step: $\kappa_{ai}\Delta t \leq 1$.\footnote{To make this point more explicit, since $\kappa_{ai}\Delta t$ is time-reparametrization invariant, it is convenient to work with proper time: $\kappa_{ai}\Delta t = \kappa_{ai}^{\rm proper} \Delta \tau_i$.
The natural proper unit time step is a Hubble time, $\Delta \tau_i = H_i^{-1}$, hence we obtain $\kappa_{ai}\Delta t = \kappa_{ai}^{\rm proper}H_i^{-1}$.
By definition, a necessary condition for~$i$ to be a vacuum is that its proper decay rate is at most its Hubble rate: $\kappa_{ai}^{\rm proper}\leq H_i$.
It follows that $\kappa_{ai}\Delta t \leq 1$.} Therefore the trapping probability satisfies the upper bound:
\be
{\cal S}_a \leq 1\,.
\label{Sa < 1}
\ee

\item It follows from the explicit expression~\eqref{lambda 1 explicit} for~$\lambda_1$ that 
\be
\sum_{a = 1}^{N_{\rm term}} {\cal S}_a = \sum_{a = 1}^{N_{\rm term}} \frac{\sum_i \kappa_{ai}\Delta t \sqrt{\omega_i}v_i^{(1)}}{\sum_i \sqrt{\omega_i}v_i^{(1)}} = |\lambda_1|\Delta t\,.
\ee

\end{enumerate}

Analogously to the relation~\eqref{SRT with terms} for dS vacua between escape probability and pMFPT, we define a pMFPT for terminals as the reciprocal
of the trapping probability:
\be
{\cal T}_a \equiv {\cal S}_a^{-1} = \frac{\sum_i \sqrt{\omega_i}v_i^{(1)}}{\sum_i \kappa_{ai}\Delta t \sqrt{\omega_i}v_i^{(1)}} \,.
\label{pMFPT terms def}
\ee 
Given~\eqref{Sa < 1}, ${\cal T}_a$ satisfies a lower bound
\be
{\cal T}_a \geq 1\,,
\label{pMFPT lower bound terms}
\ee
which is akin to~\eqref{pMFPT bound} in the dS-only case. 

\subsection{General definition of the accessibility measure}

The accessibility measure on a general landscape with non-terminal (dS) and terminal vacua 
is defined once again as the reciprocal of the pMFPT:
\be
p_I =  \frac{{\cal T}_I^{-1}}{\sum_K {\cal T}_K^{-1}} \,,
\label{p gen def}
\ee
where the MFPT is given by~\eqref{pMFPT dS def with terms} and~\eqref{pMFPT terms def} for dS vacua and terminals, respectively:
\bea
\nonumber
& & {\cal T}_i  = \sum_{\ell=2}^{N_{\rm inf}} \frac{v^{(\ell)\,2}_i}{|\lambda_\ell|\Delta t}     \qquad \qquad\;\;\;\;~ i\in {\rm dS} \,;\\
& & {\cal T}_a = \frac{\sum_i \sqrt{\omega_i}v_i^{(1)}}{\sum_i \kappa_{ai}\Delta t \sqrt{\omega_i}v_i^{(1)}}   \qquad a \in {\rm terminals} \,.
\label{pMFPT dS and terms}
\eea
As argued above, the measure thus defined is gauge invariant and independent of initial conditions. 
It is well-defined for both non-terminals and terminals alike. Because it is defined in terms of first-passage probabilities,~$p_I$ is oblivious
to the comoving {\it vs} physical volume ambiguity that afflicts stationary measures. Unlike stationary measures,~$p_I$
can be reliably calculated to leading order in the downward approximation, and therefore is insensitive to small tweaks to the
landscape. 

Importantly, the accessibility measure makes concrete and potentially testable predictions, as discussed in the next Section.

\section{Phenomenological Implications}
\label{pheno}

In this Section we derive various phenomenological implications of the accessibility measure. Importantly, these predictions do
not rely on anthropic reasoning, and instead derive from the measure itself. To be precise, in various places below we will use as
input the observed value of the cosmological constant, or equivalently, the Hubble constant~$H_0$. We do not attempt to explain the
measured~$H_0$, and its smallness may ultimately rely on anthropics. However, taking $H_0$ as a given we derive predictions
for other observables, such as the optimal lifetime of our universe and the absence of new physics at the LHC, which follow
readily from the measure.

Some of the predictions derived below, such as the lifetime of our universe, were originally obtained in~\cite{Khoury:2019yoo}.
The difference in the present analysis is that such predictions are now firmly rooted in a measure. Other predictions, such as the
optimal access time, are new.

\subsection{Access time}
\label{access time sec}

The accessibility measure~\eqref{p gen def} favors vacua that are easily accessed under time evolution, specifically
vacua with order unity pMFPT\footnote{For terminal vacua, the pMFPT satisfies a lower bound~\eqref{pMFPT lower bound terms},
hence~\eqref{pMFPT O(1)} is justified. For dS vacua, we have not yet been able to derive a strict lower bound, as discussed
around~\eqref{pMFPT dS almost lower bound}. Nevertheless, it seems reasonable to assume that~\eqref{pMFPT O(1)} holds with an order unity factor.} 
\be
{\cal T}_I \sim {\cal O}(1)\,.
\label{pMFPT O(1)}
\ee
From~\eqref{pMFPT dS and terms} we see that the pMFPT is made dimensionless by the coarse-graining time
step~$\Delta t$. Thus~\eqref{pMFPT O(1)} implies that optimal vacua are accessed in a physical time of
order~$\Delta t$. In terms of proper time for vacuum~$I$, the natural coarse-graining time interval is
course the Hubble time, $\Delta t = \Delta\tau_I = H_I^{-1}$. Therefore vacua favored by the measure are reached
in a proper time of order their own Hubble time:
\be
\tau^{\rm access}_I \sim H_I^{-1}\,.
\label{access time Hubble time}
\ee

Given the observed value of the vacuum energy in our universe,~\eqref{access time Hubble time} implies that we live
approximately $H_0^{-1} \sim 13.8$~billion years after the beginning of eternal inflation. This is not a trivial statement.
While eternal inflation is well-known to be geodesically past-incomplete~\cite{Borde:2001nh}, the last period of inflation
which gave rise to our universe could have occurred an arbitrarily long time after the initial ``big bang", in principle much
longer than $13.8$~billion years.

In particular,~\eqref{access time Hubble time} implies an upper bound on the duration of the last period of inflation. 
Denoting the Hubble scale of this last inflationary bout by $H_{\rm inf}$, the number of e-folds allowed
by the optimal access time is bounded:
\be
{\cal N} \;\lsim\; \frac{H_{\rm inf}}{H_0}\,.
\label{N bound}
\ee
Later on we will combine this with the optimal lifetime of dS vacua to derive a lower bound on $H_{\rm inf}$.

\subsection{Funnel topography}
\label{funnel sec}

To proceed, it is helpful to focus on a finite fiducial region of the landscape comprised of $N\gg 1$ vacua. For simplicity the region is
approximated as a closed system, ignoring the exchange of probability with its surroundings. We will relax this assumption below. 

Our task is to define a characteristic time for the landscape dynamics of dS vacua in the region, as a suitable average
over the pMFPTs. To start with, recall from~\eqref{kemeny with terms} the global MFPT, or Kemeny's constant, for
dS vacua in the region:
\be
{\cal T}_{\rm MFPT} = \sum_{\ell= 2}^{N_{\rm inf}} \frac{1}{|\lambda_\ell|\Delta t}\,.
\label{kemeny bis}
\ee
This gives a characteristic time for dS vacua to populate each other in the region. Meanwhile, the characteristic
time for dS vacua to populate terminals can be estimated as the reciprocal of the total escape probability for terminal vacua:
\be
{\cal T}_{\rm terms} \equiv \frac{1}{\sum_a {\cal S}_a} =  \frac{1}{|\lambda_1|\Delta t} \,.
\label{avg T terms}
\ee
Equations~\eqref{kemeny bis} and~\eqref{avg T terms} make sense. The time required to populate dS vacua is set by all but
the smallest (in magnitude) eigenvalue of the transition matrix, while the smallest eigenvalue~$\lambda_1$ sets the characteristic leakage
time into terminals. 

We are therefore led to define an average pMFPT in the region by 
\be
\langle {\cal T} \rangle \equiv \frac{{\cal T}_{\rm MFPT} + {\cal T}_{\rm terms}}{N_{\rm inf}} = \frac{1}{N_{\rm inf}} \sum_{\ell=1}^{N_{\rm inf}} \frac{1}{|\lambda_\ell|\Delta t}\,.
\ee
This gives a characteristic time for populating dS and terminal vacua in the region. Conveniently, it only depends on the eigenvalues of the transition matrix, not on its eigenvectors.
Indeed, our ulterior motive for considering a finite region of the landscape and defining an average pMFPT over that region was to achieve this simplification.

In particular, to leading order in the downward approximation,~$M_{ij}$ reduces to an upper-triangular matrix given by the first term in~\eqref{M original}, and its eigenvalues can therefore
be read off from the diagonal entries:
\be
\langle {\cal T} \rangle \simeq \frac{1}{N_{\rm inf}} \sum_{i=1}^{N_{\rm inf}} \frac{1}{\kappa_i \Delta t}  \qquad (\text{downward})\,.
\label{avg T down}
\ee
Thus~$\langle {\cal T} \rangle$ is interpreted as a dimensionless {\it mean residency time}. Incidentally, since~$\kappa_i$ is the total decay rate of node~$i$ per unit time~$t$,~\eqref{avg T down} makes the time-reparametrization invariance of~$\langle {\cal T} \rangle$ manifest. In what follows it will be convenient to work in terms of proper time $\kappa_i\Delta t = \kappa_i^{\rm proper} \Delta \tau_i$. The natural proper time step is of course the Hubble time, $\Delta\tau_i = H_i^{-1}$, thus~\eqref{avg T down} becomes
\be
\langle {\cal T} \rangle \simeq \frac{1}{N_{\rm inf}} \sum_{i=1}^{N_{\rm inf}}  H_i\tau_i^{\rm decay} \,,
\label{avg T down 2}
\ee
where $\tau_i^{\rm decay} \equiv 1/\kappa_i^{{\rm proper}}$ is the proper lifetime of dS vacuum~$i$. And since each vacuum must have by definition a proper lifetime 
longer than its Hubble time, $\tau_i^{\rm decay} \;\gsim\; H_i^{-1}$, this implies
\be
\langle {\cal T} \rangle \;\gsim\; 1\,.
\label{avg T bound}
\ee

Now we arrive at a key point. In the downward approximation it is possible for multiple dS vacua to become absolutely stable,~$\kappa_i = 0$. This will occur whenever 
such vacua have only up-tunneling as allowed transitions. In this case~\eqref{avg T down} implies that~$\langle {\cal T} \rangle$ will
diverge to leading order in the downward approximation, meaning that at sub-leading order~$\langle {\cal T} \rangle$ will be exponentially
large. A region that includes such vacua exhibits frustration and glassy dynamics~\cite{glassylandscape}. It should be clear that such
frustrated regions are heavily disfavored by the accessibility measure. 

Instead, the accessibility measure favors regions whose dS vacua all have allowed downward transitions, either to lower-lying dS vacua or to terminals. 
Such favored regions therefore have the topography of a {\it broad valley or funnel}, as sketched in Fig.~\ref{optimal region}. This is akin to the {\it principle of minimal frustration}
of protein energy landscapes~\cite{proteins1,proteins2}, where the high-energy unfolded states are connected to the lowest-energy native state by a relatively smooth funnel.
This is a key prediction of the accessibility measure. Unlike stationary measures, which rest on the idea that our vacuum should be run-of-the-mill among all hospitable vacua on the
landscape (the ``principle of mediocrity"), the accessibility measure favors vacua residing in special, funnel-like regions of the landscape. This may have important implications for 
string phenomenology and model-building.  

\subsection{Average lifetime of vacua and computational complexity}

To make further predictions, we follow~\cite{Khoury:2019yoo} and take the continuum limit of~\eqref{avg T down}, valid for~$N_{\rm inf}\gg 1$.
Given the form of CDL transitions, the lifetime of a given vacuum in general depends both on its potential energy~$V$ as well as the various
``bounce" parameters~$\theta$ characterizing the shape of the potential barrier:
\be
\tau_i^{\rm decay} = \tau_i^{\rm decay}(V_i,\theta_i)\,. 
\ee
Let~${\cal P}(V,\theta)$ denote the underlying joint probability distribution that a given vacuum has potential energy~$V$ and bounce parameters~$\theta$. For simplicity
we will assume that on the string landscape the absolute height of a vacuum and the shape of the surrounding potential barriers are
uncorrelated:~${\cal P}(V,\theta) \equiv {\cal P}(V) \hat{{\cal P}}(\theta)$. This allows us to marginalize over bounce parameters and define an average
lifetime~$\tau_{\rm decay}(V)$ for vacua of given potential energy:
\be
\tau_{\rm decay}(V) \equiv \int {\rm d}\theta \,\tau_{\rm decay}(V,\theta)\hat{{\cal P}}(\theta)\,.
\ee
Therefore, using the Friedmann relation $H \sim\sqrt{V}/M_{\rm Pl}$, the mean residency time~\eqref{avg T down 2}
becomes, in the continuum limit,
\be
\langle {\cal T} \rangle =  \int_{V_{\rm min}}^{V_{\rm max}} {\rm d}V  \frac{\sqrt{V}}{M_{\rm Pl}}\, \tau_{\rm decay}(V) \,{\cal P}(V) \,,
\label{avg T cont}
\ee
where $V_{\rm min}$ and $V_{\rm max}$ are respectively the smallest and largest vacuum energy achieved in the region.

Provided that~${\cal P}(V)$ falls off sufficiently fast at large~$V$, the result for~$\langle {\cal T} \rangle$ is controlled by the
behavior of~$\tau_{\rm decay}(V)$ for small~$V$. Assuming as usual that ${\cal P}(V)$ is nearly uniform for~$V$ much smaller
than the fundamental scale~\cite{Weinberg:2000qm},~\eqref{avg T cont} can be approximated by
\be
\langle {\cal T} \rangle \sim \int_{V_{\rm min}} {\rm d}V  \frac{\sqrt{V}}{M_{\rm Pl}}\, \tau_{\rm decay}(V) \,,
\label{avg T cont 2}
\ee
where, for a uniform distribution, the smallest potential energy is on average inversely proportional to the number of vacua:
\be
V_{\rm min} \sim \frac{M_{\rm Pl}^4}{N_{\rm inf}}\,.
\ee
Clearly the integral~\eqref{avg T cont 2} will converge or diverge as $V_{\rm min}\rightarrow 0$ depending on whether 
$\tau_{\rm decay}(V)$ diverges slower or faster than~$V^{-3/2}$, with the critical case~$\tau_{\rm decay}(V) \sim V^{-3/2}$ corresponding to a
logarithmic divergence. Note that, since vacua must have a lifetime longer than their Hubble time, at the very least we have
$\tau_{\rm decay}(V) > M_{\rm Pl}/\sqrt{V}$. 

The accessibility measure favors regions of the landscape where the mean residency time~$\langle {\cal T} \rangle$ nearly
saturates~\eqref{avg T bound}. This requires the integral~\eqref{avg T cont 2} to converge or, at worst,
depend logarithmically on $V_{\rm min}$. Using the Planck mass $M_{\rm Pl}$ to fix dimensions, as it is the
natural scale in the problem, the measure favors landscape regions where vacua have an average lifetime in the
range
\be
\frac{M_{\rm Pl}}{\sqrt{V}} < \tau_{\rm decay}(V) \;\lsim\; \frac{M_{\rm Pl}^5}{V^{3/2}}\qquad \text{as}~~V \rightarrow 0\,.
\label{lifetime range}
\ee
The shortest allowed lifetime is of course the Hubble time. The longest allowed lifetime is recognized as the Page time~\cite{Page:1993wv}
for dS space~\cite{Danielsson:2002td,Danielsson:2003wb,Ferreira:2016hee,Ferreira:2017ogo}:
\be
\tau_{\rm Page} \sim  \frac{M_{\rm Pl}^5}{V^{3/2}} \sim \frac{M_{\rm Pl}^2}{H^3}\,.
\label{page time} 
\ee
In slow-roll inflation, the Page time marks the phase transition to slow-roll eternal inflation~\cite{Creminelli:2008es} and has been
used to place a bound on the maximum number of e-folds that can be described semi-classically~\cite{ArkaniHamed:2007ky}. The
appearance of the Page time in the present context of false-vacuum eternal inflation, first noticed in~\cite{Khoury:2019yoo}, is surprising. 

From a computational complexity perspective, the Page time represents a transition in the scaling of~$\langle {\cal T} \rangle$ with the
number of vacua. To see this, let us assume for simplicity that $\tau_{\rm decay}(V)$ is a power-law for small~$V$, 
\be
\tau_{\rm decay}(V) \sim V^{-\alpha} \qquad \text{as}~~V \rightarrow 0\,.
\ee
Then the mean residency time~\eqref{avg T cont 2} gives
\be
\langle {\cal T} \rangle  \sim \left\{\begin{array}{ccl}
\text{constant}  & ~~\text{for} & \alpha < 3/2  \\
\log V_{\rm min}\sim \log N_{\rm inf} & ~~\text{for} &\alpha = 3/2  \\
V_{\rm min}^{3/2 - \alpha} \sim N_{\rm inf}^{\alpha-3/2} & ~~\text{for} & \alpha > 3/2 \,.
\end{array}\right.
\label{comp scaling}
\ee
In turn, the number of vacua~$N_{\rm inf}$ generically scales exponentially with the effective moduli-space dimensionality~$D$ of the landscape region,
that is,~$N_{\rm inf} \sim {\rm e}^D$. We learn from~\eqref{comp scaling} that regions with slow transition rates,~$\alpha > 3/2$, correspond to
a mean residency time scaling polynomially in~$N_{\rm inf}$, hence exponentially in~$D$. This is compatible with the~\textsf{NP}-hard complexity class
of finding vacua within a suitable range of potential energy~\cite{Denef:2006ad}. Regions with fast enough transition rates,~$\alpha \leq 3/2$, on the other hand, have a mean residency
time scaling at worst logarithmically in~$N_{\rm inf}$, hence linearly in $D$. Note that this does not contradict the~\textsf{NP}-hard complexity classification,
as \textsf{NP}-hardness, being a worst-case assessment, does not preclude the existence of polynomial-time solutions for special instances of the problem.

The case $\alpha = 3/2$, where the average lifetime is order the Page time, marks a critical boundary between the other two phases.
The mean residency time diverges as $\log N_{\rm inf}$, signaling a {\it dynamical phase transition}.  A similar non-equilibrium phase transition occurs in quenched disordered media, whenever the probability distribution for waiting times reaches a critical power-law~\cite{disordered media}. It also describes a {\it computational phase transition}~\cite{compPT1,compPT2}.
A famous example is the phase transition in heuristic decision-tree pruning from polynomial to exponential search time at a critical value of the effective branching ratio~\cite{pruning}.

\subsection{Recurrence and dynamical criticality}
\label{criticality page}

The lifetime range~\eqref{lifetime range} preferred by the accessibility measure was derived by approximating the finite landscape region of interest
as a closed system. With this assumption, minimizing $\langle {\cal T} \rangle$ requires that downward transitions are as fast as possible. More realistically, 
however, one should treat regions as open systems, allowing for the possibility that a random walker escapes a given region 
before accessing a target vacuum. It stands to reason that vacua residing in regions where the likelihood of escape is high should be disfavored
by the measure. We expect that the accessibility measure favors regions where random walks efficiently explore vacua, 
thereby minimizing the mean residency time, while at the same time minimizing the likelihood of escape before finding viable vacua.

Treating landscape regions as open systems would require modeling the probability leaking into environment, which may introduce
unwanted model-dependence in our analysis. Following~\cite{Khoury:2019yoo}, we instead propose to study a proxy requirement that relies
solely on the intrinsic dynamics within a given region. Specifically, we demand that random walks in the region are {\it recurrent}
in the infinite-network limit,~$N_{\rm inf} \rightarrow \infty$. As discussed in Sec.~\ref{escape}, in recurrent walks every site in the region 
will be visited with probability one. Recurrent walks thoroughly explore any region around their starting point,
whereas transient walks tend to escape to infinity. Although not formally equivalent to modeling regions as open systems,
recurrence offers a reliable and model-independent benchmark for efficient sampling~\cite{Khoury:2019yoo}. 

Per~\eqref{recur relation S} random walks will be recurrent at~$i$ when~$N_{\rm inf} \rightarrow \infty$ if the escape
probability~\eqref{escape prob with terms}~${\cal S}_i$ vanishes in this limit. In turn, from~\eqref{SRT with terms} this
requires that the pMFPT~${\cal T}_i$ diverges in the limit. Therefore, random walks in a given region will be recurrent if
the mean residency time diverges as~$N_{\rm inf} \rightarrow \infty$:
\be
\langle {\cal T} \rangle \rightarrow \infty \qquad \text{as}~~N_{\rm inf}\rightarrow\infty\,.
\label{Tau kappa}
\ee

We therefore have two competing requirements: minimal mean residency time, which requires that vacua have relatively short lifetimes, per~\eqref{lifetime range};
and recurrence, which requires that the mean residency time diverge as $N_{\rm inf}\rightarrow\infty$. Optimal regions reach a compromise by achieving the
shortest $\langle {\cal T} \rangle$ compatible with recurrence, {\it i.e.}, the least-divergent integral~\eqref{avg T cont 2}. Per~\eqref{comp scaling}, 
vacua in optimal regions have an average lifetime of order the Page time~\eqref{page time}:
\be
\tau_{\rm crit} (H) \sim \frac{M_{\rm Pl}^2}{H^3}\,.
\label{tau crit}
\ee
That the Page time represents an optimal time for vacuum selection on the landscape was first realized in~\cite{Khoury:2019yoo}.
In the present analysis this is now justified by considerations of a well-defined measure. 

As discussed earlier, the average lifetime of order the Page time corresponds to a dynamical phase transition. 
Thus the joint demands of minimal oversampling, defined by minimal mean residency time, and sweeping exploration,
defined by recurrence, {\it selects regions of the landscape that are tuned at criticality.} Therefore, we are led to conjecture that
the accessibility measure favors dynamically critical regions of the landscape. We cannot yet claim a rigorous proof of this statement,
because recurrence is only a proxy for minimizing the escape probability, but it is reasonable to expect that the accessibility measure,
being rooted in search optimization, is peaked at criticality.

Complex self-organized systems poised at criticality are ubiquitous in the natural world~\cite{living}. Examples include brain activity, 
where the probability distribution for neuronal avalanches of different size is scale invariant~\cite{brain 1,brain 2,brain 3,brain 4};
and the flocking behavior of starlings, whose velocity correlations are scale invariant~\cite{flock obs,flock dynamics}. It has been conjectured
that dynamical criticality is evolutionarily favored because it offers an ideal compromise between robust response to external stimuli
and flexibility of adaptation to a changing environment.

Furthermore, it has been argued that computational capabilities are maximized at the phase transition between stable and unstable
dynamical behavior --- the so-called ``edge of chaos"~\cite{dynamical crit review}. This idea goes back to random boolean networks~\cite{Kauffman} and cellular
automata~\cite{wolfram,edge of chaos 1,edge of chaos 2,edge of chaos 3,edge of chaos 4,edge of chaos 5}. In machine learning,
certain recurrent neural networks~\cite{RNNchaos1,RNNchaos2} achieve maximal computational power for vanishing Lyapunov
exponent~\cite{RNNchaos3}. Meanwhile, the connectivity matrix of well-trained, state-of-the-art deep neural networks has been
shown recently to have a power-law spectral density~\cite{DNNpowerlaw}, well-described by heavy-tailed random matrix theory~\cite{RMT}.

Similarly, our mechanism selects regions of the landscape that are dynamically critical, in the sense of the recurrence/transience
dynamical phase transition, and computationally critical, in the sense that~$\langle {\cal T} \rangle$ lies at the transition between
polynomial and non-polynomial search time. Tantalizingly, this suggests a connection between non-equilibrium critical phenomena
on the landscape and the near-criticality of our universe. We illustrate this below with Higgs metastability.

\subsection{Higgs metastability and particle phenomenology}

If our vacuum is part of an optimal region of the landscape, characterized by vacua
with critical Page lifetimes~\eqref{tau crit}, then we predict that the lifetime of our universe is
\be
\tau \sim \frac{M_{\rm Pl}^2}{H_0^3} \sim 10^{130}~{\rm years}\,.
\label{tdecay pred}
\ee
This explains the metastability of the electroweak vacuum. Remarkably, the predicted lifetime agrees
to within~$\gsim\; 2\sigma$ with the SM prediction~\cite{Andreassen:2017rzq}:~$\tau_{\rm SM} = 10^{526^{+409}_{-202}}~{\rm years}$.
To be clear, we of course do not claim to explain the smallness of the cosmological constant. But taking the observed vacuum energy $\sim M_{\rm Pl}^2 H_0^2$ as given,
the optimal lifetime~\eqref{tdecay pred} constrains a combination of SM parameters, including the Higgs and top quark masses. Indeed, what makes
Higgs metastability particularly interesting is the relation it entails between the cosmological constant and weak hierarchy problems.

Closer agreement with the SM lifetime estimate can be achieved if the top quark is slightly heavier, $m_{\rm t} \simeq 174.5~{\rm GeV}$.
This can be viewed as a prediction, assuming of course  that the SM is valid all the way to the Planck scale. New physics at intermediate scales can reduce the tension. For instance,
adding a gauge-invariant, higher-dimensional operator $h^6/\Lambda_{\rm NP}^2$ will affect the predicted lifetime
if $\Lambda_{\rm NP}\;\lsim\; 10^{13}~{\rm GeV}$, assuming the central value $m_{\rm t} = 173.5~{\rm GeV}$~\cite{Andreassen:2017rzq}. 

More generally, our mechanism offers a dynamical explanation for why our universe is poised at criticality. It gives a {\it raison d'\^{e}tre} for the conspiracy
underlying Higgs metastability. In other words, from the point of view of the accessibility measure the inferred metastability of the electroweak vacuum
is sacred. New physics below the SM instability scale, $\sim 10^{10}~{\rm GeV}$, on the other hand, can jeopardize this
observable. Here are some of the implications for a few BSM candidates already outlined in~\cite{Khoury:2019yoo}: 

\begin{itemize}

\item {\it Low-scale SUSY:} If the SUSY breaking scale is $\lsim\;10^{10}~{\rm GeV}$, this will directly impact the stability of our vacuum. There are three obvious possibilities: 1)~SUSY makes our vacuum unstable ({\it e.g.}, via decay to charge/color breaking vacua~\cite{Gunion:1987qv}), which by itself is inconsistent and therefore requires additional new physics; 2)~SUSY makes our vacuum stable, which is disfavored by our mechanism; 3)~SUSY maintains our vacuum within the metastability region. The latter possibility, while logically consistent with our mechanism, would require further numerical conspiracy, above and beyond that already achieved in the SM. Barring fine-tunings, the natural implication of SUSY below $10^{10}~{\rm GeV}$ is to make our vacuum stable,\footnote{This expectation is borne out by an explicit calculation of~\cite{Giudice:2011cg}, which showed that if all SUSY partners have masses at the SUSY breaking scale,
then the metastability of our vacuum requires a SUSY breaking scale of $\gsim\;10^{10}~{\rm GeV}$.} which is disfavored by the measure. 

Therefore the accessibility measure favors optimal regions of the landscape characterized by very high-scale SUSY breaking. More generally, the above argument applies 
to any new physics at the LHC. It follows that {\it the discovery of BSM particles at the LHC, including low-scale SUSY, would rule out the possibility that our vacuum lies
in an optimal region of the landscape.} This is a falsifiable prediction of the accessibility measure.

\item {\it Sterile neutrinos:} Massive right-handed neutrinos, like the top quark, tend to make the vacuum less stable. Assuming three right-handed neutrinos of comparable mass, for simplicity, the impact on Higgs metastability is negligible if their mass is $\lsim\; 10^{13}~{\rm GeV}$~\cite{EliasMiro:2011aa}. On the other hand, if their mass is around $10^{13}-10^{14}~{\rm GeV}$, then the expected lifetime for our vacuum will be in closer agreement with the optimal lifetime~\eqref{tdecay pred}.  

\item {\it QCD axion:} Consider the QCD axion as a solution to the strong CP problem. The radial part of the~$U(1)$ complex scalar is a boson and hence makes the electroweak vacuum more stable. To preserve the desired metastability, the Peccei-Quinn scale must be sufficiently high, $f_{\rm a} \;\gsim\; 10^{10}~{\rm GeV}$~\cite{Hertzberg:2012zc}.

\end{itemize}

% Think about thermal dark matter and Higgs metastability...

\subsection{Scale of inflation and amplitude of gravitational waves}

Vacua in optimal regions of the landscape are, on the one hand, accessed in a Hubble time, per~\eqref{access time Hubble time}, and, on the other,
have a lifetime of order the Page time~\eqref{tau crit}. This implies a bound on the scale of the last period of inflation for our universe.

Assuming that our vacuum is part of an optimal region of the landscape, then the parent dS vacuum which tunneled to our vacuum had a proper
lifetime of order its Page time, $\tau_{\rm parent} \sim M_{\rm Pl}^2/H_{\rm parent}^3$, corresponding to a number of e-folds of   
\be
{\cal N}_{\rm parent} \sim \frac{M_{\rm Pl}^2}{H_{\rm parent}^2} \,.
\ee
But since our vacuum was accessed within a time~$H_0^{-1}$, the number of e-folds is bounded by~\eqref{N bound}:
\be
{\cal N}_{\rm parent}\;\lsim\; \frac{H_{\rm parent}}{H_0}\,.
\ee
It follows that
\be
H_{\rm parent} \;\gsim\; \left(M_{\rm Pl}^2 H_0\right)^{1/3} \simeq 20~{\rm MeV}\,. 
\label{H parent bound}
\ee

It is usually assumed that the tunneling event from the parent vacuum is followed by a period of slow-roll inflation,
with Hubble scale~$H_{\rm inf}$. Assuming $H_{\rm inf} \sim H_{\rm parent}$ for concreteness, then~\eqref{H parent bound}
implies a lower bound on the slow-roll inflationary energy scale:
\be
E_{\rm inf} \sim \sqrt{H_{\rm parent}M_{\rm Pl}} \;\gsim\; 10^8~{\rm GeV} \,.
\label{Einf lower}
\ee
Therefore the accessibility measure disfavors low-scale inflationary scenarios. 
 
On the other hand, it has been argued that if the inflationary is too high, then Higgs quantum fluctuations during inflation
can push the field beyond the potential barrier~\cite{Espinosa:2015qea}. Assuming minimal coupling of the Higgs to gravity,
for simplicity, the inflationary scale must satisfy~$H_{\rm inf}\;\lsim\; 10^{9}~{\rm GeV}$, or~$E_{\rm inf}\;\lsim\; 10^{14}~{\rm GeV}$, 
to keep Higgs fluctuations under control. (The bound becomes looser with non-minimal coupling of suitable sign~\cite{Espinosa:2015qea} and can be affected by higher-dimensional operators and deviations from exact de Sitter~\cite{Fumagalli:2019ohr}.) Combined with~\eqref{Einf lower}, we conclude that the optimal range for the inflationary scale, assuming a minimally-coupled Higgs, is
\be
10^8~{\rm GeV}\;\lsim\; E_{\rm inf}\;\lsim\; 10^{14}~{\rm GeV} \,.
\ee
A detection of primordial gravitational waves, for instance from cosmic microwave background polarization observations, would imply that the Higgs must have a non-minimal coupling to gravity, or otherwise disfavor the possibility that our vacuum lies in an optimal region of the landscape. 

\section{Conclusions}

The measure problem is arguably the most pressing and formidable challenge in cosmology. If the fundamental theory allows eternal inflation,
then our universe is but a small region of a vast multiverse containing infinitely-many other pocket universes with different physical properties. A
measure is therefore necessary to make {\it any} predictions about physical observables in our universe.

Two broad classes of measures have been proposed: i)~global measures, that count pocket universes on a global foliation of the eternally-inflating
space-time; ii)~local measures, that count pocket universes in a finite region of space-time defined by a single observer. Each approach has pros
and cons. Global measures are independent of initial conditions but depend sensitively on the choice of foliation as well as whether bubbles are
weighted according to comoving or physical volume. Local measures are manifestly gauge-invariant but are sensitive to initial conditions. 

A drawback afflicting both global and local measures is their sensitivity to minor tweaks of the landscape. Global and local measures are based on the nearly-stationary distribution of the Markov process describing vacuum dynamics. The stationary distribution exponentially favors a single dS vacuum --- the dominant vacuum, which has the slowest decay rate anywhere on the landscape. This {\it is} the robust prediction of global/local measures. But since this so-called dominant vacuum is unlikely to be hospitable, one is forced to keep track of
subleading terms in the transition matrix, which encode upward transitions from the dominant vacuum to hospitable vacua. As shown in the Appendix,
exponentially small tweaks in the transition rates can result in exponential differences in the relative probabilities for hospitable vacua. 

Thus, despite more than three decades of effort, the measure problem remains unsolved. However, two aspects of the problem give us hope that
a solution is within reach. Firstly, after suitable coarse-graining the rate equation governing vacuum dynamics reduces to a simple, linear
Markov process, free of the conceptual pitfalls of eternal inflation. Secondly, despite the variety of approaches to the measure problem, all
proposed measures to date have focused on a single statistics: the stationary distribution of the Markov process. Granted, this is the most
natural and simplest statistics to consider. But, as the large body of recent work on complex networks has shown, the stationary distribution
offers only a narrow viewpoint of the importance of different nodes. Specifically, it is predominantly sensitive to the local properties of the network.
This can be seen most emphatically in the dS-only case, where $f^\infty_i  =  \frac{w_i}{w}$ is determined solely by the vacuum energy of each
node and conveys no information about network topology. It is oblivious, in particular, to whether a given node is well-connected or
isolated from other vacua.

The purpose of this work is to offer a fresh approach to the measure problem, drawing on recent results in network science. 
Indeed, from the broader perspective of random walks on graphs, the measure problem translates to a question of network centrality. 
Various centrality measures have been proposed in the literature, each offering different perspectives on the importance
of nodes in a network. Some measures, such as degree centrality, are determined by local properties of the network, akin to the stationary
measure on the landscape. Other centrality indices offer insights on dynamical aspects of the network, by identifying nodes that control
information flow.

The accessibility measure presented in this paper belongs to this latter category. Instead of characterizing the distribution of vacua near equilibrium, the proposed
measure pertains to the approach to equilibrium. It favors vacua that are easily accessed and populated early on in the evolution. Specifically, such
vacua are populated within a time of order their Hubble time, much earlier than the exponentially long mixing time for the landscape. This is motivated
physically by the possibility that the eternal inflation has been unfolding for a time much shorter than the relaxation time. 

As such the accessibility measure naturally connects to issues of computational complexity and search optimization on the landscape. Generic regions of the landscape are characterized by frustrated dynamics, resulting in exponentially long search times compatible with the~\textsf{NP}-hard complexity class of the general problem. In contrast, the
accessibility measure favors regions of the landscape where the search algorithm is efficient. Such optimal regions can be thought of as
special, polynomial-time instances of the general problem.

The proposed measure enjoys a number of desirable features. It is simultaneously time-reparametrization invariant, independent
of initial conditions, and oblivious to whether pocket universes are weighted according to their comoving or physical volume. It does not suffer
from youngness bias or Boltzmann brains. Unlike stationary measures, it is robust against minor tweaks to the landscape. It can be
reliably calculated to leading order in the downward approximation, {\it i.e.}, neglecting the exponentially small terms in the transition matrix
that encode upward transitions.

Importantly, the accessibility measure makes concrete, falsifiable predictions that are largely independent of anthropic reasoning. 
The most enticing prediction is that our vacuum should have a lifetime of order its Page time, $\sim 10^{130}$~years. 
Remarkably, this agrees to within~$\gsim\; 2\sigma$ with the SM result for electroweak metastability. To be precise,
the predicted lifetime takes $H_0$ as an input (perhaps anthropically determined) and then constrains a combination of the SM
parameters that are most important in determining vacuum stability, namely the Higgs mass, top quark mass, and gauge couplings. Therefore,
given a (possibly anthropic) solution to the cosmological constant problem, the accessibility measure offers a non-anthropic solution to the
weak hierarchy problem. The appearance of the Page time as a critical time for landscape dynamics is somewhat mysterious, and it will be interesting
to seek a deeper understanding for its origin in this context.

Thus search optimization on the landscape offers a compelling explanation for the delicate numerical conspiracy underlying Higgs metastability. 
From this point of view the inferred metastability of the electroweak vacuum is no accident --- it is a sacred observable. Any new BSM physics
discovered at the LHC, including low-scale SUSY, would have to conspire to maintain the metastability bound, which would require
further fine-tuning above and beyond that already achieved in the SM. Therefore, barring fine tuning, it stands to reason that the discovery of BSM particles at the LHC or future colliders,
including low-scale SUSY, would rule out the possibility that our vacuum lies in an optimal region of the landscape. This still allows the
possibility of discovering light particles, such as the axion, as these would have negligible impact on vacuum stability. These are falsifiable predictions of the accessibility measure. 

Another key prediction of the accessibility measure is that our vacuum lies a special region of the landscape with funnel-like
topography. This is in stark contrast with the ``principle of mediocrity" underlying stationary measures, whereby our vacuum
should be run-of-the-mill among all hospitable vacua on the landscape. The most interesting implications of funnel-like regions may be for string phenomenology and model-building. String vacua
with realistic particle physics are usually considered in isolation, without much consideration for their accessibility and the topography
of the surrounding landscape region. It will be very interesting to study the implications of a funnel topography on particle physics, in particular whether SM-like particle
spectra are more prevalent in such regions. Relatedly, as already speculated in~\cite{Khoury:2019yoo}, it would be interesting to see
whether funnel-like regions of the landscape correspond to a low effective moduli-space dimensionality, particularly in the vicinity
of the lowest-energy vacuum, as this could offer a dynamical explanation for why our universe does not have more than three
spatial dimensions (though anthropic reasoning may be necessary to explain why it does not have fewer than three).

A third key prediction of the accessibility measure is that it favors regions of the landscape tuned at dynamical criticality.
This suggests a connection between criticality of vacuum dynamics on the landscape and the near-criticality of our universe,
as the vacuum metastability prediction already illustrates. Indeed, it is striking that most fine-tuning problems in fundamental physics,
such as the weak hierarchy problem and the cosmological constant problem, can also be understood as problems of criticality. 
It is tempting to speculate that search optimization might shed new, non-anthropic light on the smallness of the cosmological constant.
Last but not least is slow-roll inflation, which itself represents a phenomenon of near-criticality as the inflaton interpolates between
an approximately conformally invariant de Sitter phase and standard big bang cosmology. It will be very interesting to see
whether slow-roll potentials are more ubiquitous in funnel-like regions of the landscape than in random places in moduli space.

\vspace{.4cm}
\noindent
{\bf Acknowledgements:} We thank Cliff Burgess, Jaume Garriga, Alan Guth for helpful discussions, and are particularly grateful to James Halverson,
Cody Long, Onkar Parrikar and Alex Vilenkin for their insights. We thank the Institut de Ci\`encies del Cosmos at the University of Barcelona (ICCUB) for their hospitality while part of this work was completed. This work is supported by the US Department of Energy (HEP) Award DE-SC0013528, NASA ATP grant 80NSSC18K0694, and by the Simons Foundation Origins of the Universe Initiative.

\appendix

\section*{Appendix: Sensitivity of Stationary Measures}
\label{sensitivity}

In this Appendix we point out a drawback of local and global measures based on the nearly-stationary distribution of the Markov process --- their sensitivity to exponentially small terms in the transition matrix. The root of the problem is that the stationary distribution overwhelmingly favors a single dS vacuum --- the one with the slowest decay rate. Because this so-called dominant vacuum is unlikely to be hospitable, the relative probabilities for different hospitable vacua are determined by upward transitions from the dominant vacuum, which in turn are sensitive to exponentially small terms in the transition matrix. Therefore, exponentially small tweaks in the transition rates can result in exponential differences in the relative probabilities for hospitable vacua. While not logistically inconsistent, we view such sensitivity to minor tweaks to the landscape as undesirable. After giving the general argument in more detail below, we will illustrate this with a toy mini-landscape comprised of a three dS vacua.

As we have seen in~\eqref{f stationary terms}, whenever the landscape includes terminal vacua the stationary solution $f^\infty$ lies entirely
in the terminal subspace. The relative probability to lie in different dS vacua is then determined by the subleading term as $t\rightarrow \infty$,
which in turn is set by the eigenvalue of~$\mathbb{M}$ with the largest (non-vanishing) real part. Per our discussion below~\eqref{lamba's M}, this
``dominant" eigenvalue is just $\lambda_1$, and the corresponding eigenvector is $v^{(1)}_{\mathbb{M}}$. Therefore at late times we have
\be
f(t) \simeq f^\infty + \beta \,v^{(1)}_{\mathbb{M}} {\rm e}^{\lambda_1 t}\,,
\label{f subleading}
\ee
with $\beta$ a constant. Thus, since $v^{(\ell)}_{\mathbb{M}\,i} = v^{(\ell)}_{M\,i}$ per~\eqref{vbig M}, the
relative probability for different dS vacua is set by the leading eigenvector of $M$.

In the downward approximation, recall from~\eqref{downward lambda 1} that $\lambda_1$ is set by the smallest decay rate. Correspondingly,
the dominant eigenvector $v^{(1)}_{M\,i}$ is set by the {\it longest-lived dS vacuum} --- the so-called dominant vacuum. While the detailed
nature of this vacuum requires input from string theory, on general grounds one expects that it has very small vacuum energy and is
surrounded by vacua of much higher potential energy. In this configuration, its only allowed CDL transitions involve ``upward" tunneling, such that
its decay rate is exponentially suppressed (per~\eqref{detailed balance}).

{\it A priori} there is no reason to expect that the dominant vacuum is hospitable. Thus the relative probabilities of different
hospitable dS vacua is set by their relative transition rates from the dominant vacuum. And because such transitions necessarily involve up-tunneling,
as argued above, the relative probabilities are sensitive to the exponentially small contributions to the transition matrix encoded in~$M_{\rm up}$ in~\eqref{M original}. 
It follows that {\it exponentially small changes to the transition matrix can result in dramatically different predictions for the relative abundance of hospitable vacua.}
As we will see in the toy example below, small tweaks to the landscape can have a major impact on~$M_{\rm up}$, and consequently on the predictions of the
stationary measure. Furthermore, these tweaks can be done in such a way as to preserve all downward transition rates, as well as the hierarchy of
potential energies between different dS vacua. 

To be clear, the stationary measure {\it does} make a robust prediction. Among all dS vacua, it overwhelmingly favors a single one --- the longest-lived
dS vacuum. Indeed, the dominant eigenvector appearing in~\eqref{f subleading} can be accurately determined in the downward approximation, as
argued above, and as such is insensitive to exponentially small corrections encoded in $M_{\rm up}$. The problem, of course, is that, unless
we are exceedingly lucky, the dominant vacuum will be inhospitable. In this case the relative probabilities for hospitable vacua are controlled by
exponentially small upward transition rates, hence the sensitivity to small tweaks to the landscape.
 
\begin{figure}[htb]
\centering
\includegraphics[scale=0.6]{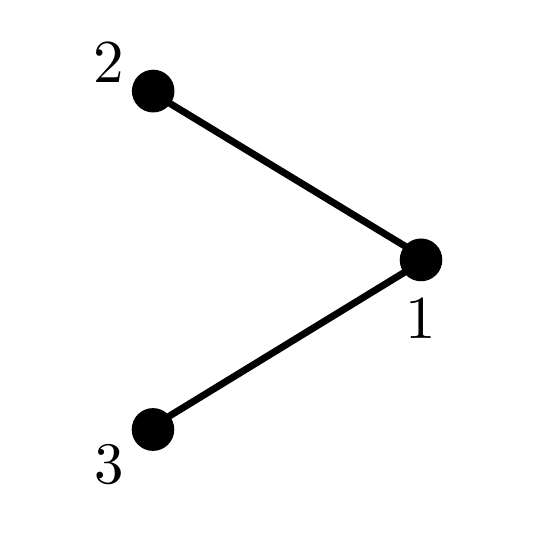}
\caption{A toy mini-landscape comprised of vacuum~1, the dominant vacuum, which is connected to vacua~2 and~3 with much higher potential energy. The mini-landscape also includes terminal vacua, not shown in the Figure.}  
\label{mini}
\end{figure} 
 
To illustrate these general points, consider a simple mini-landscape comprised of three dS vacua (Fig.~\ref{mini}). The mini-landscape also includes terminals,
not shown in the Figure. Vacuum~1, the dominant vacuum, is connected to only two other dS vacua labeled by~2 and~3, both assumed hospitable. The assumed
hierarchy of potential energies is
\be
0 < V_1 \ll V_2, V_3\,,
\label{Vhierarchy2}
\ee
while we remain agnostic for the moment about the relative magnitude of~$V_2$ and~$V_3$.

We are therefore interested in determining the relative probability~$p_2/p_3$ to occupy vacuum~2 or~3. In global measures where vacua are weighted
according to volume,~$p_2/p_3$ is given directly by $f_2/f_3$. In local measures, such as the watcher measure, the relative probability
is instead given by the relative number of bubbles,~$N_2/N_3$, encountered along the watcher's time-like geodesic~\cite{Garriga:2005av}. In either case, the result is 
set by the transition rate from the dominant vacuum to the hospitable ones: 
\be
\frac{p_2}{p_3} \sim \frac{\kappa_{21}}{\kappa_{31}} = \frac{w_2}{w_3}\,\frac{\kappa_{12}}{\kappa_{13}}\,,
\label{f2f3 2}
\ee
where the last step follows from detailed balance~\eqref{detailed balance}. The upshot is that the ratio $p_2/p_3$ is
sensitive to the exponentially small upward transition rates~$\kappa_{21}$ and~$\kappa_{31}$.

In particular, it is possible to change the ratio $p_2/p_3$ while keeping all downward rates fixed. Indeed, imagine varying
$V_3$ while keeping $V_2$, as well as the downward rates $\kappa_{12}$ and $\kappa_{13}$, fixed.
(Since the latter is given $\kappa_{13} = A_{13}/w_3$, this entails adjusting the shape of the potential such that $A_{13}$ and $w_3$ vary in
tandem.) In the process of letting $V_3 \rightarrow \hat{V}_3$ the relative fraction~\eqref{f2f3 2} changes to a new value $\widehat{p_2/p_3}$ given by
\be
\frac{\widehat{p_2/p_3}}{p_2/p_3} = \frac{w_3}{\hat{w}_3} \sim {\rm e}^{48\pi^2M_{\rm Pl}^4\left(\frac{1}{V_3}-\frac{1}{\hat{V}_3}\right)}\,.
\ee
Clearly this ratio can be $\ll 1$ or $\gg 1$ depending on whether $V_3$ is larger or smaller than $\hat{V}_3$. For instance, the relative fraction can change
from $p_2/p_3 \ll 1$ (vacuum 3 heavily disfavored) to $\widehat{p_2/p_3}\gg 1$ (vacuum 3 heavily favored), thereby inverting
the predicted volume fraction of type 2 {\it vs} type 3 regions. Note that this not require changing the hierarchy between $V_2$ and $V_3$.
For instance, suppose that $V_2 \ll V_3$, then with the above procedure it is possible, for suitable downward rates, to dramatically change
the relative probability of vacuum~2 and~3 while maintaining the hierarchy of potential energies. 

It is straightforward to concoct more elaborate examples, but the basic point remains the same --- the predictions of stationary
measures for hospitable vacua are sensitive to exponentially small terms in the transition matrix. Minor tweaks to the landscape
can alter these exponentially small corrections, resulting in dramatically different predictions. 

In contrast, the accessibility measure presented in the main text is based on first-passage statistics, which in turn are determined by
the eigenspectrum of the transition matrix. The eigenvalues and eigenvectors of the matrix~\eqref{M original} are to an
excellent approximation determined by the leading upper-triangular matrix, encoding downward transitions, up to exponentially small corrections
due to~$M_{\rm up}$. Small tweaks to the landscape, as in the toy example above, have negligible impact on the accessibility measure.
In this sense measures based on first-passage statistics are robust against small changes to the landscape.


\begin{thebibliography}{99}

%\cite{Vilenkin:1983xq}
\bibitem{Vilenkin:1983xq} 
  A.~Vilenkin,
  ``The Birth of Inflationary Universes,''
  Phys.\ Rev.\ D {\bf 27}, 2848 (1983).
%  doi:10.1103/PhysRevD.27.2848
  %%CITATION = doi:10.1103/PhysRevD.27.2848;%%

%\cite{Linde:1986fc}
\bibitem{Linde:1986fc} 
  A.~D.~Linde,
  ``Eternal Chaotic Inflation,''
  Mod.\ Phys.\ Lett.\ A {\bf 1}, 81 (1986).
 % doi:10.1142/S0217732386000129
  %%CITATION = doi:10.1142/S0217732386000129;%%

%\cite{Linde:1986fd}
\bibitem{Linde:1986fd} 
  A.~D.~Linde,
  ``Eternally Existing Selfreproducing Chaotic Inflationary Universe,''
  Phys.\ Lett.\ B {\bf 175}, 395 (1986).
%  doi:10.1016/0370-2693(86)90611-8
  %%CITATION = doi:10.1016/0370-2693(86)90611-8;%%

%\cite{Freivogel:2011eg}
\bibitem{Freivogel:2011eg} 
  B.~Freivogel,
  ``Making predictions in the multiverse,''
  Class.\ Quant.\ Grav.\  {\bf 28}, 204007 (2011)
 % doi:10.1088/0264-9381/28/20/204007
  [arXiv:1105.0244 [hep-th]].
  %%CITATION = doi:10.1088/0264-9381/28/20/204007;%%

%\cite{Bousso:2000xa}
\bibitem{Bousso:2000xa} 
  R.~Bousso and J.~Polchinski,
  ``Quantization of four form fluxes and dynamical neutralization of the cosmological constant,''
  JHEP {\bf 0006}, 006 (2000)
%  doi:10.1088/1126-6708/2000/06/006
  [hep-th/0004134].
  %%CITATION = doi:10.1088/1126-6708/2000/06/006;%%

%\cite{Kachru:2003aw}
\bibitem{Kachru:2003aw} 
  S.~Kachru, R.~Kallosh, A.~D.~Linde and S.~P.~Trivedi,
  ``De Sitter vacua in string theory,''
  Phys.\ Rev.\ D {\bf 68}, 046005 (2003)
%  doi:10.1103/PhysRevD.68.046005
  [hep-th/0301240].
  %%CITATION = doi:10.1103/PhysRevD.68.046005;%%

%\cite{Susskind:2003kw}
\bibitem{Susskind:2003kw} 
  L.~Susskind,
  ``The Anthropic landscape of string theory,''
  In *Carr, Bernard (ed.): Universe or multiverse?* 247-266
  [hep-th/0302219].
  %%CITATION = HEP-TH/0302219;%%

%\cite{Douglas:2003um}
\bibitem{Douglas:2003um} 
  M.~R.~Douglas,
  ``The Statistics of string / M theory vacua,''
  JHEP {\bf 0305}, 046 (2003)
%  doi:10.1088/1126-6708/2003/05/046
  [hep-th/0303194].
  %%CITATION = doi:10.1088/1126-6708/2003/05/046;%%

%\cite{Obied:2018sgi}
\bibitem{Obied:2018sgi} 
  G.~Obied, H.~Ooguri, L.~Spodyneiko and C.~Vafa,
  ``De Sitter Space and the Swampland,''
  arXiv:1806.08362 [hep-th].
  %%CITATION = ARXIV:1806.08362;%%

%\cite{Agrawal:2018own}
\bibitem{Agrawal:2018own} 
  P.~Agrawal, G.~Obied, P.~J.~Steinhardt and C.~Vafa,
  ``On the Cosmological Implications of the String Swampland,''
  Phys.\ Lett.\ B {\bf 784}, 271 (2018)
%  doi:10.1016/j.physletb.2018.07.040
  [arXiv:1806.09718 [hep-th]].
  %%CITATION = doi:10.1016/j.physletb.2018.07.040;%%

%\cite{Garg:2018reu}
\bibitem{Garg:2018reu} 
  S.~K.~Garg and C.~Krishnan,
  ``Bounds on Slow Roll and the de Sitter Swampland,''
  JHEP {\bf 1911}, 075 (2019)
%  doi:10.1007/JHEP11(2019)075
  [arXiv:1807.05193 [hep-th]].
  %%CITATION = doi:10.1007/JHEP11(2019)075;%%
 
%\cite{Ooguri:2018wrx}
\bibitem{Ooguri:2018wrx} 
  H.~Ooguri, E.~Palti, G.~Shiu and C.~Vafa,
  ``Distance and de Sitter Conjectures on the Swampland,''
  Phys.\ Lett.\ B {\bf 788}, 180 (2019)
%  doi:10.1016/j.physletb.2018.11.018
  [arXiv:1810.05506 [hep-th]].
  %%CITATION = doi:10.1016/j.physletb.2018.11.018;%%

%\cite{Palti:2019pca}
\bibitem{Palti:2019pca} 
  E.~Palti,
  ``The Swampland: Introduction and Review,''
  Fortsch.\ Phys.\  {\bf 67}, no. 6, 1900037 (2019)
%  doi:10.1002/prop.201900037
  [arXiv:1903.06239 [hep-th]].
  %%CITATION = doi:10.1002/prop.201900037;%%

%\cite{Linde:1993nz}
\bibitem{Linde:1993nz} 
  A.~D.~Linde and A.~Mezhlumian,
  ``Stationary universe,''
  Phys.\ Lett.\ B {\bf 307}, 25 (1993)
%  doi:10.1016/0370-2693(93)90187-M
  [gr-qc/9304015].
  %%CITATION = doi:10.1016/0370-2693(93)90187-M;%%

%\cite{Linde:1993xx}
\bibitem{Linde:1993xx} 
  A.~D.~Linde, D.~A.~Linde and A.~Mezhlumian,
  ``From the Big Bang theory to the theory of a stationary universe,''
  Phys.\ Rev.\ D {\bf 49}, 1783 (1994)
%  doi:10.1103/PhysRevD.49.1783
  [gr-qc/9306035].
  %%CITATION = doi:10.1103/PhysRevD.49.1783;%%

%\cite{GarciaBellido:1993wn}
\bibitem{GarciaBellido:1993wn} 
  J.~Garcia-Bellido, A.~D.~Linde and D.~A.~Linde,
  ``Fluctuations of the gravitational constant in the inflationary Brans-Dicke cosmology,''
  Phys.\ Rev.\ D {\bf 50}, 730 (1994)
%  doi:10.1103/PhysRevD.50.730
  [astro-ph/9312039].
  %%CITATION = doi:10.1103/PhysRevD.50.730;%%

%\cite{Vilenkin:1994ua}
\bibitem{Vilenkin:1994ua} 
  A.~Vilenkin,
  ``Predictions from quantum cosmology,''
  Phys.\ Rev.\ Lett.\  {\bf 74}, 846 (1995)
%  doi:10.1103/PhysRevLett.74.846
  [gr-qc/9406010].
  %%CITATION = doi:10.1103/PhysRevLett.74.846;%%
 
%\cite{DeSimone:2008if}
\bibitem{DeSimone:2008if} 
  A.~De Simone, A.~H.~Guth, A.~D.~Linde, M.~Noorbala, M.~P.~Salem and A.~Vilenkin,
  ``Boltzmann brains and the scale-factor cutoff measure of the multiverse,''
  Phys.\ Rev.\ D {\bf 82}, 063520 (2010)
 % doi:10.1103/PhysRevD.82.063520
  [arXiv:0808.3778 [hep-th]].
  %%CITATION = doi:10.1103/PhysRevD.82.063520;%%
  
 %\cite{Bousso:2008hz}
\bibitem{Bousso:2008hz} 
  R.~Bousso, B.~Freivogel and I.~S.~Yang,
  ``Properties of the scale factor measure,''
  Phys.\ Rev.\ D {\bf 79}, 063513 (2009)
%  doi:10.1103/PhysRevD.79.063513
  [arXiv:0808.3770 [hep-th]].
  %%CITATION = doi:10.1103/PhysRevD.79.063513;%%  
  
%\cite{DeSimone:2008bq}
\bibitem{DeSimone:2008bq} 
  A.~De Simone, A.~H.~Guth, M.~P.~Salem and A.~Vilenkin,
  ``Predicting the cosmological constant with the scale-factor cutoff measure,''
  Phys.\ Rev.\ D {\bf 78}, 063520 (2008)
%  doi:10.1103/PhysRevD.78.063520
  [arXiv:0805.2173 [hep-th]].
  %%CITATION = doi:10.1103/PhysRevD.78.063520;%%  
   
%\cite{Guth:2007ng}
\bibitem{Guth:2007ng} 
  A.~H.~Guth,
  ``Eternal inflation and its implications,''
  J.\ Phys.\ A {\bf 40}, 6811 (2007)
%  doi:10.1088/1751-8113/40/25/S25
  [hep-th/0702178 [HEP-TH]].
  %%CITATION = doi:10.1088/1751-8113/40/25/S25;%%   
  
%\cite{Vilenkin:2019mwc}
\bibitem{Vilenkin:2019mwc} 
  A.~Vilenkin and M.~Yamada,
  ``Four-volume cutoff measure of the multiverse,''
  Phys.\ Rev.\ D {\bf 101}, no. 4, 043520 (2020)
 % doi:10.1103/PhysRevD.101.043520
  [arXiv:1912.02187 [hep-th]].
  %%CITATION = doi:10.1103/PhysRevD.101.043520;%%  
  
 %\cite{Garriga:1997ef}
\bibitem{Garriga:1997ef} 
  J.~Garriga and A.~Vilenkin,
  ``Recycling universe,''
  Phys.\ Rev.\ D {\bf 57}, 2230 (1998)
 % doi:10.1103/PhysRevD.57.2230
  [astro-ph/9707292].
  %%CITATION = doi:10.1103/PhysRevD.57.2230;% 
   
%\cite{Garriga:2005av}
\bibitem{Garriga:2005av} 
  J.~Garriga, D.~Schwartz-Perlov, A.~Vilenkin and S.~Winitzki,
  ``Probabilities in the inflationary multiverse,''
  JCAP {\bf 0601}, 017 (2006)
 % doi:10.1088/1475-7516/2006/01/017
  [hep-th/0509184].
  %%CITATION = doi:10.1088/1475-7516/2006/01/017;%%  

%\cite{Vanchurin:2006qp}
\bibitem{Vanchurin:2006qp} 
  V.~Vanchurin and A.~Vilenkin,
  ``Eternal observers and bubble abundances in the landscape,''
  Phys.\ Rev.\ D {\bf 74}, 043520 (2006)
%  doi:10.1103/PhysRevD.74.043520
  [hep-th/0605015].
  %%CITATION = doi:10.1103/PhysRevD.74.043520;%%

 %\cite{Bousso:2006ev}
\bibitem{Bousso:2006ev} 
  R.~Bousso,
  ``Holographic probabilities in eternal inflation,''
  Phys.\ Rev.\ Lett.\  {\bf 97}, 191302 (2006)
%  doi:10.1103/PhysRevLett.97.191302
  [hep-th/0605263].
  %%CITATION = doi:10.1103/PhysRevLett.97.191302;%%

%\cite{Bousso:2009dm}
\bibitem{Bousso:2009dm} 
  R.~Bousso,
  ``Complementarity in the Multiverse,''
  Phys.\ Rev.\ D {\bf 79}, 123524 (2009)
 % doi:10.1103/PhysRevD.79.123524
  [arXiv:0901.4806 [hep-th]].
  %%CITATION = doi:10.1103/PhysRevD.79.123524;%%

%\cite{Bousso:2010zi}
\bibitem{Bousso:2010zi} 
  R.~Bousso, B.~Freivogel, S.~Leichenauer and V.~Rosenhaus,
  ``A geometric solution to the coincidence problem, and the size of the landscape as the origin of hierarchy,''
  Phys.\ Rev.\ Lett.\  {\bf 106}, 101301 (2011)
%  doi:10.1103/PhysRevLett.106.101301
  [arXiv:1011.0714 [hep-th]].
  %%CITATION = doi:10.1103/PhysRevLett.106.101301;%%


%\cite{Garriga:2012bc}
\bibitem{Garriga:2012bc} 
  J.~Garriga and A.~Vilenkin,
  ``Watchers of the multiverse,''
  JCAP {\bf 1305}, 037 (2013)
%  doi:10.1088/1475-7516/2013/05/037
  [arXiv:1210.7540 [hep-th]].
  %%CITATION = doi:10.1088/1475-7516/2013/05/037;%%  

%\cite{Nomura:2011dt}
\bibitem{Nomura:2011dt} 
  Y.~Nomura,
  ``Physical Theories, Eternal Inflation, and Quantum Universe,''
  JHEP {\bf 1111}, 063 (2011)
 % doi:10.1007/JHEP11(2011)063
  [arXiv:1104.2324 [hep-th]].
  %%CITATION = doi:10.1007/JHEP11(2011)063;%%  

%\cite{Garriga:2013cix}
\bibitem{Garriga:2013cix} 
  J.~Garriga, A.~Vilenkin and J.~Zhang,
  ``Non-singular bounce transitions in the multiverse,''
  JCAP {\bf 1311}, 055 (2013)
%  doi:10.1088/1475-7516/2013/11/055
  [arXiv:1309.2847 [hep-th]].
  %%CITATION = doi:10.1088/1475-7516/2013/11/055;%%  
     
 %\cite{Bousso:2009mw}
\bibitem{Bousso:2009mw} 
  R.~Bousso and I.~S.~Yang,
  ``Global-Local Duality in Eternal Inflation,''
  Phys.\ Rev.\ D {\bf 80}, 124024 (2009)
  doi:10.1103/PhysRevD.80.124024
  [arXiv:0904.2386 [hep-th]].
  %%CITATION = doi:10.1103/PhysRevD.80.124024;%%
  
\bibitem{centrality review}
L. Da F.~Costa, F.~A.~Rodrigues, G.~Travieso and P.~R.~Villas Boas,
``Characterization of complex networks: A survey of measurements,"
 Advances in Phys. {\bf 56}, 167 (2007)
 [arXiv:cond-mat/0505185].
  
\bibitem{Katz}
L.~Katz,
``A new status index derived from sociometric analysis,"
Psychometrika {\bf 18}, 39 (1953).  
  
\bibitem{pagerank}
S.~Brin and L.~Page,
``The anatomy of a large-scale hypertextual Web search engine,"
Computer Networks and ISDN Systems {\bf 30}, 107 (1998).

\bibitem{closeness1}
A.~Bavelas,
``Communication patterns in task-oriented groups,"
J. Acoust. Soc. Am, {\bf 22}, 725 (1950).

\bibitem{closeness2}
G.~Sabidussi, 
``The centrality index of a graph,"
Psychometrika {\bf 31}, 581 (1966).

%\cite{Borde:2001nh}
\bibitem{Borde:2001nh} 
  A.~Borde, A.~H.~Guth and A.~Vilenkin,
  ``Inflationary space-times are incomplete in past directions,''
  Phys.\ Rev.\ Lett.\  {\bf 90}, 151301 (2003)
%  doi:10.1103/PhysRevLett.90.151301
  [gr-qc/0110012].
  %%CITATION = doi:10.1103/PhysRevLett.90.151301;%%

%\cite{Denef:2011ee}
\bibitem{Denef:2011ee} 
  F.~Denef,
  ``TASI lectures on complex structures,''
 % doi:10.1142/9789814350525_0007
  arXiv:1104.0254 [hep-th].
  %%CITATION = doi:10.1142/9789814350525_0007;%%  

%\cite{Denef:2017cxt}
\bibitem{Denef:2017cxt} 
  F.~Denef, M.~R.~Douglas, B.~Greene and C.~Zukowski,
  ``Computational complexity of the landscape II --- Cosmological considerations,''
  Annals Phys.\  {\bf 392}, 93 (2018)
%  doi:10.1016/j.aop.2018.03.013
  [arXiv:1706.06430 [hep-th]].
  %%CITATION = doi:10.1016/j.aop.2018.03.013;%%

%\cite{Khoury:2019yoo}
\bibitem{Khoury:2019yoo} 
  J.~Khoury and O.~Parrikar,
  ``Search Optimization, Funnel Topography, and Dynamical Criticality on the String Landscape,''
  JCAP {\bf 1912}, 014 (2019)
  [arXiv:1907.07693 [hep-th]].


%%\cite{Linde:2007nm}
%\bibitem{Linde:2007nm} 
%  A.~D.~Linde,
%  ``Towards a gauge invariant volume-weighted probability measure for eternal inflation,''
%  JCAP {\bf 0706}, 017 (2007)
%%  doi:10.1088/1475-7516/2007/06/017
%  [arXiv:0705.1160 [hep-th]].
%  %%CITATION = doi:10.1088/1475-7516/2007/06/017;%%
%  
%%\cite{Linde:2008xf}
%\bibitem{Linde:2008xf} 
%  A.~D.~Linde, V.~Vanchurin and S.~Winitzki,
%  ``Stationary Measure in the Multiverse,''
%  JCAP {\bf 0901}, 031 (2009)
%%  doi:10.1088/1475-7516/2009/01/031
%  [arXiv:0812.0005 [hep-th]].
%  %%CITATION = doi:10.1088/1475-7516/2009/01/031;%%  

\bibitem{funnelfig}
A.~Samarakoon {\it et al.},
``Aging, memory, and nonhierarchical energy landscape of spin jam,"
Proc. Natl Acad. Sci USA {\bf 113}, 11806 (2016)
[arXiv:1707.03086 [cond-mat.dis-nn]].   
 
\bibitem{proteins1}
J.~D.~Bryngelson, J.~N.~Onuchic, N.~D.~Socci and P.~G.~Wolynes,
``Funnels, Pathways and the Energy Landscape of Protein Folding: A Synthesis,"
Proteins-Struct. Func. and Genetics. {\bf 21}, 167 (1995)
[chem-ph/9411008].
 
\bibitem{proteins2}
N.~G$\bar{{\rm o}}$,
``Theoretical Studies of Protein Folding,"
Ann. Rev. Biophys. Bioeng. {\bf 12}, 183 (1983).
  
\bibitem{MFPT book}
S.~Redner,
``A Guide to First-Passage Processes,"
Cambridge University Press, 328 p. (2001). 
  
\bibitem{GMFPT}
V.~Tejedor, O.~B\'enichou and R.~Voituriez,
``Global mean first-passage times of random walks on complex networks,"
Phys. Rev. E {\bf 80}, 065104(R) (2009)
[arXiv:0909.0657 [cond-mat.stat-mech]].  
  
\bibitem{complex network PRL}
J.~D.~Noh and H.~Rieger,
``Random Walks on Complex Networks,"
 Phys.\ Rev.\ Lett.\  {\bf 92}, 118701 (2004).
  
\bibitem{2ndorder}
A.-M.~Kermarrec, E.~Le Merrer, B.~Sericola and G.~Tr\'edan,
``Second order centrality: Distributed assessment of nodes criticity in complex networks,"  
Computer Comm. {\bf 34}, 619 (2011). 
  
\bibitem{glassylandscape}
J.~C.~Mauro and M.~M.~Smedskjaer,
``Statistical mechanics of glass,"
Journal of Non-Crystalline Solids {\bf 396}, 41 (2014). 
  
%\cite{Andreassen:2017rzq}
\bibitem{Andreassen:2017rzq} 
  A.~Andreassen, W.~Frost and M.~D.~Schwartz,
  ``Scale Invariant Instantons and the Complete Lifetime of the Standard Model,''
  Phys.\ Rev.\ D {\bf 97}, no. 5, 056006 (2018)
 % doi:10.1103/PhysRevD.97.056006
  [arXiv:1707.08124 [hep-ph]].
  %%CITATION = doi:10.1103/PhysRevD.97.056006;%%
   
%\cite{EliasMiro:2011aa}
\bibitem{EliasMiro:2011aa} 
  J.~Elias-Miro, J.~R.~Espinosa, G.~F.~Giudice, G.~Isidori, A.~Riotto and A.~Strumia,
  ``Higgs mass implications on the stability of the electroweak vacuum,''
  Phys.\ Lett.\ B {\bf 709}, 222 (2012)
%  doi:10.1016/j.physletb.2012.02.013
  [arXiv:1112.3022 [hep-ph]].
  %%CITATION = doi:10.1016/j.physletb.2012.02.013;%%
 
%\cite{Espinosa:2015qea}
\bibitem{Espinosa:2015qea} 
  J.~R.~Espinosa, G.~F.~Giudice, E.~Morgante, A.~Riotto, L.~Senatore, A.~Strumia and N.~Tetradis,
  ``The cosmological Higgstory of the vacuum instability,''
  JHEP {\bf 1509}, 174 (2015)
%  doi:10.1007/JHEP09(2015)174
  [arXiv:1505.04825 [hep-ph]].
  %%CITATION = doi:10.1007/JHEP09(2015)174;%%

%\cite{Fumagalli:2019ohr}
\bibitem{Fumagalli:2019ohr} 
  J.~Fumagalli, S.~Renaux-Petel and J.~W.~Ronayne,
  ``Higgs vacuum (in)stability during inflation: the dangerous relevance of de Sitter departure and Planck-suppressed operators,''
  arXiv:1910.13430 [hep-ph].
  %%CITATION = ARXIV:1910.13430;%%
      
%\cite{Denef:2006ad}
\bibitem{Denef:2006ad} 
  F.~Denef and M.~R.~Douglas,
  ``Computational complexity of the landscape. I.,''
  Annals Phys.\  {\bf 322}, 1096 (2007)
%  doi:10.1016/j.aop.2006.07.013
  [hep-th/0602072].
  %%CITATION = doi:10.1016/j.aop.2006.07.013;%%
 
 %\cite{ArkaniHamed:2005yv}
\bibitem{ArkaniHamed:2005yv} 
  N.~Arkani-Hamed, S.~Dimopoulos and S.~Kachru,
  ``Predictive landscapes and new physics at a TeV,''
  hep-th/0501082.
  %%CITATION = HEP-TH/0501082;%%
 
%\cite{Bao:2017thx}
\bibitem{Bao:2017thx} 
  N.~Bao, R.~Bousso, S.~Jordan and B.~Lackey,
  ``Fast optimization algorithms and the cosmological constant,''
  Phys.\ Rev.\ D {\bf 96}, no. 10, 103512 (2017)
%  doi:10.1103/PhysRevD.96.103512
  [arXiv:1706.08503 [hep-th]].
  %%CITATION = doi:10.1103/PhysRevD.96.103512;%%
 
%\cite{Halverson:2018cio}
\bibitem{Halverson:2018cio} 
  J.~Halverson and F.~Ruehle,
  ``Computational Complexity of Vacua and Near-Vacua in Field and String Theory,''
  Phys.\ Rev.\ D {\bf 99}, no. 4, 046015 (2019)
%  doi:10.1103/PhysRevD.99.046015
  [arXiv:1809.08279 [hep-th]].
  %%CITATION = doi:10.1103/PhysRevD.99.046015;%%
 
%\cite{Cvetic:2010ky}
\bibitem{Cvetic:2010ky} 
  M.~Cvetic, I.~Garcia-Etxebarria and J.~Halverson,
  ``On the computation of non-perturbative effective potentials in the string theory landscape: IIB/F-theory perspective,''
  Fortsch.\ Phys.\  {\bf 59}, 243 (2011)
%  doi:10.1002/prop.201000093
  [arXiv:1009.5386 [hep-th]].
  %%CITATION = doi:10.1002/prop.201000093;%%
  
%\cite{Halverson:2019vmd}
\bibitem{Halverson:2019vmd} 
  J.~Halverson, M.~Plesser, F.~Ruehle and J.~Tian,
  ``Kahler Moduli Stabilization and the Propagation of Decidability,''
  arXiv:1911.07835 [hep-th].
  %%CITATION = ARXIV:1911.07835;%% 
 
 
%% COSMOLOGICAL 
%\cite{Carifio:2017nyb}
\bibitem{Carifio:2017nyb} 
  J.~Carifio, W.~J.~Cunningham, J.~Halverson, D.~Krioukov, C.~Long and B.~D.~Nelson,
  ``Vacuum Selection from Cosmology on Networks of String Geometries,''
  Phys.\ Rev.\ Lett.\  {\bf 121}, no. 10, 101602 (2018)
%  doi:10.1103/PhysRevLett.121.101602
  [arXiv:1711.06685 [hep-th]].
  %%CITATION = doi:10.1103/PhysRevLett.121.101602;%%  

   
%\cite{He:2017aed}
\bibitem{He:2017aed} 
  Y.~H.~He,
  ``Deep-Learning the Landscape,''
  arXiv:1706.02714 [hep-th].
  %%CITATION = ARXIV:1706.02714;%%

 %\cite{Krefl:2017yox}
\bibitem{Krefl:2017yox} 
  D.~Krefl and R.~K.~Seong,
  ``Machine Learning of Calabi-Yau Volumes,''
  Phys.\ Rev.\ D {\bf 96}, no. 6, 066014 (2017)
%  doi:10.1103/PhysRevD.96.066014
  [arXiv:1706.03346 [hep-th]].
  %%CITATION = doi:10.1103/PhysRevD.96.066014;%% 
 
%\cite{Ruehle:2017mzq}
\bibitem{Ruehle:2017mzq} 
  F.~Ruehle,
  ``Evolving neural networks with genetic algorithms to study the String Landscape,''
  JHEP {\bf 1708}, 038 (2017)
%  doi:10.1007/JHEP08(2017)038
  [arXiv:1706.07024 [hep-th]].
  %%CITATION = doi:10.1007/JHEP08(2017)038;%% 
 
%\cite{Carifio:2017bov}
\bibitem{Carifio:2017bov} 
  J.~Carifio, J.~Halverson, D.~Krioukov and B.~D.~Nelson,
  ``Machine Learning in the String Landscape,''
  JHEP {\bf 1709}, 157 (2017)
%  doi:10.1007/JHEP09(2017)157
  [arXiv:1707.00655 [hep-th]].
  %%CITATION = doi:10.1007/JHEP09(2017)157;%% 
  
%\cite{Wang:2018rkk}
\bibitem{Wang:2018rkk} 
  Y.~N.~Wang and Z.~Zhang,
  ``Learning non-Higgsable gauge groups in 4D F-theory,''
  JHEP {\bf 1808}, 009 (2018)
%  doi:10.1007/JHEP08(2018)009
  [arXiv:1804.07296 [hep-th]].
  %%CITATION = doi:10.1007/JHEP08(2018)009;%% 

%\cite{Klaewer:2018sfl}
\bibitem{Klaewer:2018sfl} 
  D.~Klaewer and L.~Schlechter,
  ``Machine Learning Line Bundle Cohomologies of Hypersurfaces in Toric Varieties,''
  Phys.\ Lett.\ B {\bf 789}, 438 (2019)
%  doi:10.1016/j.physletb.2019.01.002
  [arXiv:1809.02547 [hep-th]].
  %%CITATION = doi:10.1016/j.physletb.2019.01.002;%% 
  
%\cite{Mutter:2018sra}
\bibitem{Mutter:2018sra} 
  A.~Mütter, E.~Parr and P.~K.~S.~Vaudrevange,
  ``Deep learning in the heterotic orbifold landscape,''
  Nucl.\ Phys.\ B {\bf 940}, 113 (2019)
  doi:10.1016/j.nuclphysb.2019.01.013
  [arXiv:1811.05993 [hep-th]].
  %%CITATION = doi:10.1016/j.nuclphysb.2019.01.013;%%
 
%\cite{Cole:2018emh}
\bibitem{Cole:2018emh} 
  A.~Cole and G.~Shiu,
  ``Topological Data Analysis for the String Landscape,''
  JHEP {\bf 1903}, 054 (2019)
%  doi:10.1007/JHEP03(2019)054
  [arXiv:1812.06960 [hep-th]].
  %%CITATION = doi:10.1007/JHEP03(2019)054;%% 
 
%\cite{Halverson:2019tkf}
\bibitem{Halverson:2019tkf} 
  J.~Halverson, B.~Nelson and F.~Ruehle,
  ``Branes with Brains: Exploring String Vacua with Deep Reinforcement Learning,''
  JHEP {\bf 1906}, 003 (2019)
%  doi:10.1007/JHEP06(2019)003
  [arXiv:1903.11616 [hep-th]].
  %%CITATION = doi:10.1007/JHEP06(2019)003;%% 
 
%\cite{He:2019vsj}
\bibitem{He:2019vsj} 
  Y.~H.~He and S.~J.~Lee,
  ``Distinguishing elliptic fibrations with AI,''
  Phys.\ Lett.\ B {\bf 798}, 134889 (2019)
%  doi:10.1016/j.physletb.2019.134889
  [arXiv:1904.08530 [hep-th]].
  %%CITATION = doi:10.1016/j.physletb.2019.134889;%% 

%\cite{Cole:2019enn}
\bibitem{Cole:2019enn} 
  A.~Cole, A.~Schachner and G.~Shiu,
  ``Searching the Landscape of Flux Vacua with Genetic Algorithms,''
  JHEP {\bf 1911}, 045 (2019)
%  doi:10.1007/JHEP11(2019)045
  [arXiv:1907.10072 [hep-th]].
  %%CITATION = doi:10.1007/JHEP11(2019)045;%% 
 
\bibitem{disordered media}
S.~Havlin and D.~Ben-Avraham,
``Diffusion in disordered media,"
Advances in Physics {\bf 36}, 695 (1987).
 
%\cite{Coleman:1977py}
\bibitem{Coleman:1977py} 
  S.~R.~Coleman,
  ``The Fate of the False Vacuum. 1. Semiclassical Theory,''
  Phys.\ Rev.\ D {\bf 15}, 2929 (1977)
  Erratum: [Phys.\ Rev.\ D {\bf 16}, 1248 (1977)].
%  doi:10.1103/PhysRevD.15.2929, 10.1103/PhysRevD.16.1248
  %%CITATION = doi:10.1103/PhysRevD.15.2929, 10.1103/PhysRevD.16.1248;%%

%\cite{Callan:1977pt}
\bibitem{Callan:1977pt} 
  C.~G.~Callan, Jr. and S.~R.~Coleman,
  ``The Fate of the False Vacuum. 2. First Quantum Corrections,''
  Phys.\ Rev.\ D {\bf 16}, 1762 (1977).

\bibitem{Coleman:1980aw} 
  S.~R.~Coleman and F.~De Luccia,
  ``Gravitational Effects on and of Vacuum Decay,''
  Phys.\ Rev.\ D {\bf 21}, 3305 (1980).
  
%\cite{Lee:1987qc}
\bibitem{Lee:1987qc} 
  K.~M.~Lee and E.~J.~Weinberg,
  ``Decay of the True Vacuum in Curved Space-time,''
  Phys.\ Rev.\ D {\bf 36}, 1088 (1987).

%\cite{SchwartzPerlov:2006hi}
\bibitem{SchwartzPerlov:2006hi} 
  D.~Schwartz-Perlov and A.~Vilenkin,
  ``Probabilities in the Bousso-Polchinski multiverse,''
  JCAP {\bf 0606}, 010 (2006)
%  doi:10.1088/1475-7516/2006/06/010
  [hep-th/0601162].
   
%\cite{Olum:2007yk}
\bibitem{Olum:2007yk} 
  K.~D.~Olum and D.~Schwartz-Perlov,
  ``Anthropic prediction in a large toy landscape,''
  JCAP {\bf 0710}, 010 (2007)
%  doi:10.1088/1475-7516/2007/10/010
  [arXiv:0705.2562 [hep-th]].

\bibitem{MFPT ref}
V.~Tejedor, O.~B\'enichou, and R.~Voituriez,
``Global mean first-passage times of random walks on complex networks,"
Phys. Rev. E {\bf 80}, 065104(R) (2009)
[arXiv:0909.0657 [cond-mat.stat-mech]].

%\cite{Vennin:2015hra}
\bibitem{Vennin:2015hra} 
  V.~Vennin and A.~A.~Starobinsky,
  ``Correlation Functions in Stochastic Inflation,''
  Eur.\ Phys.\ J.\ C {\bf 75}, 413 (2015)
 % doi:10.1140/epjc/s10052-015-3643-y
  [arXiv:1506.04732 [hep-th]].

%\cite{Assadullahi:2016gkk}
\bibitem{Assadullahi:2016gkk} 
  H.~Assadullahi, H.~Firouzjahi, M.~Noorbala, V.~Vennin and D.~Wands,
  ``Multiple Fields in Stochastic Inflation,''
  JCAP {\bf 1606}, 043 (2016)
 % doi:10.1088/1475-7516/2016/06/043
  [arXiv:1604.04502 [hep-th]].
  %%CITATION = doi:10.1088/1475-7516/2016/06/043;%%

%\cite{Vennin:2016wnk}
\bibitem{Vennin:2016wnk} 
  V.~Vennin, H.~Assadullahi, H.~Firouzjahi, M.~Noorbala and D.~Wands,
  ``Critical Number of Fields in Stochastic Inflation,''
  Phys.\ Rev.\ Lett.\  {\bf 118}, no. 3, 031301 (2017)
%  doi:10.1103/PhysRevLett.118.031301
  [arXiv:1604.06017 [astro-ph.CO]].
  %%CITATION = doi:10.1103/PhysRevLett.118.031301;%%

%\cite{Noorbala:2018zlv}
\bibitem{Noorbala:2018zlv} 
  M.~Noorbala, V.~Vennin, H.~Assadullahi, H.~Firouzjahi and D.~Wands,
  ``Tunneling in Stochastic Inflation,''
  JCAP {\bf 1809}, 032 (2018)
 % doi:10.1088/1475-7516/2018/09/032
  [arXiv:1806.09634 [hep-th]].
  %%CITATION = doi:10.1088/1475-7516/2018/09/032;%%

\bibitem{Kac}
M.~Kac,
``On the notion of recurrence in discrete stochastic processes," 
Bull. Amer. Math. Soc. {\bf 53}, 1002 (1947). 

\bibitem{kemeny}
J.~G.~Kemeny and J.~L.~Snell, 
``Finite Markov Chains," 
Van Nostrand Comp. Int., New York, 1960.  
  
\bibitem{pseudogreen}
S.~Condamin {\it et al.},
``First-passage times in complex scale-invariant media,"
Nature {\bf 450}, 77 (2007).

\bibitem{Michelitsch:2017}
T.~M.~Michelitsch {\it et al.}.
``Recurrence of random walks with long-range steps generated by fractional Laplacian matrices on regular networks and simple cubic lattices,"
Journal of Phys. A: Math. and Theo. {\bf 50}, 50, 505004 (2017)
 [arXiv:1707.05843 [cond-mat.stat-mech]]. 

%\cite{Weinberg:2000qm}
\bibitem{Weinberg:2000qm} 
  S.~Weinberg,
  ``A Priori probability distribution of the cosmological constant,''
  Phys.\ Rev.\ D {\bf 61}, 103505 (2000)
%  doi:10.1103/PhysRevD.61.103505
  [astro-ph/0002387].
  %%CITATION = doi:10.1103/PhysRevD.61.103505;%%

%\cite{Page:1993wv}
\bibitem{Page:1993wv} 
  D.~N.~Page,
  ``Information in black hole radiation,''
  Phys.\ Rev.\ Lett.\  {\bf 71}, 3743 (1993)
%  doi:10.1103/PhysRevLett.71.3743
  [hep-th/9306083].
  %%CITATION = doi:10.1103/PhysRevLett.71.3743;%%

%\cite{Danielsson:2002td}
\bibitem{Danielsson:2002td} 
  U.~H.~Danielsson, D.~Domert and M.~E.~Olsson,
  ``Miracles and complementarity in de Sitter space,''
  Phys.\ Rev.\ D {\bf 68}, 083508 (2003)
%  doi:10.1103/PhysRevD.68.083508
  [hep-th/0210198].
  %%CITATION = doi:10.1103/PhysRevD.68.083508;%%

%\cite{Danielsson:2003wb}
\bibitem{Danielsson:2003wb} 
  U.~H.~Danielsson and M.~E.~Olsson,
  ``On thermalization in de Sitter space,''
  JHEP {\bf 0403}, 036 (2004)
%  doi:10.1088/1126-6708/2004/03/036
  [hep-th/0309163].
  %%CITATION = doi:10.1088/1126-6708/2004/03/036;%%

%\cite{Ferreira:2016hee}
\bibitem{Ferreira:2016hee} 
  R.~Z.~Ferreira, M.~Sandora and M.~S.~Sloth,
  ``Asymptotic Symmetries in de Sitter and Inflationary Spacetimes,''
  JCAP {\bf 1704}, no. 04, 033 (2017)
%  doi:10.1088/1475-7516/2017/04/033
  [arXiv:1609.06318 [hep-th]].
  %%CITATION = doi:10.1088/1475-7516/2017/04/033;%%

%\cite{Ferreira:2017ogo}
\bibitem{Ferreira:2017ogo} 
  R.~Z.~Ferreira, M.~Sandora and M.~S.~Sloth,
  ``Patient Observers and Non-perturbative Infrared Dynamics in Inflation,''
  JCAP {\bf 1802}, no. 02, 055 (2018)
 % doi:10.1088/1475-7516/2018/02/055
  [arXiv:1703.10162 [hep-th]].
  %%CITATION = doi:10.1088/1475-7516/2018/02/055;%%

%\cite{Creminelli:2008es}
\bibitem{Creminelli:2008es} 
  P.~Creminelli, S.~Dubovsky, A.~Nicolis, L.~Senatore and M.~Zaldarriaga,
  ``The Phase Transition to Slow-roll Eternal Inflation,''
  JHEP {\bf 0809}, 036 (2008)
%  doi:10.1088/1126-6708/2008/09/036
  [arXiv:0802.1067 [hep-th]].
  %%CITATION = doi:10.1088/1126-6708/2008/09/036;%%
      
%\cite{ArkaniHamed:2007ky}
\bibitem{ArkaniHamed:2007ky} 
  N.~Arkani-Hamed, S.~Dubovsky, A.~Nicolis, E.~Trincherini and G.~Villadoro,
  ``A Measure of de Sitter entropy and eternal inflation,''
  JHEP {\bf 0705}, 055 (2007)
 % doi:10.1088/1126-6708/2007/05/055
  [arXiv:0704.1814 [hep-th]].
  %%CITATION = doi:10.1088/1126-6708/2007/05/055;%%

\bibitem{compPT1}
R.~Monasson {\it et al.},
``Determining computational complexity from characteristic `phase transitions',"
Nature {\bf 400}, 133 (1999).

\bibitem{compPT2}
T.~Hogg, B.~A.~Huberman and C.~P.~Williams,
``Phase transitions and the search problem,"
Artificial Intelligence {\bf 81}, 1 (1996).

\bibitem{pruning}
B.~A.~Huberman and T.~Hogg, 
``Phase transitions in artificial intelligence systems," 
Artificial Intelligence {\bf 33}, 155 (1987).

\bibitem{living}
T.~Mora and W.~Bialek,
``Are biological systems poised at criticality?"
J. Stat. Phys. {\bf 144}, 268 (2011)
[arXiv:1012.2242 [q-bio.QM]].

\bibitem{brain 1}
M.~Usher, M.~Stemmler and Z.~Olami, 
``Dynamic pattern formation leads to $1/f$ noise in neural populations,"
Phys Rev Lett {\bf 74}, 326 (1995).

\bibitem{brain 2}  
O.~Kinouchi and M,~Copelli,
``Optimal Dynamical Range of Excitable Networks at Criticality,"
Nature Phys. {\bf 2}, 348 (2006)
[arXiv:q-bio/0601037 [q-bio.NC]].

\bibitem{brain 3}  
J.~M.~Beggs and D.~Plenz,
``Neuronal avalanches in neocortical circuits,"
J. Neurosci {\bf 23}, 11167 (2003).

\bibitem{brain 4}  
D.~R.~Chialvo,
``Emergent complex neural dynamics,"
Nature Phys. {\bf 6}, 744 (2010)
[arXiv:1010.2530 [q-bio.NC]].

\bibitem{flock obs}
A.~Cavagna {\it et al.}, 
``Scale-free correlations in starling flocks,"
Proc. Natl Acad. Sci USA {\bf 107}, 11865 (2010).

\bibitem{flock dynamics}
W.~Bialek {\it et al.}, 
``Statistical mechanics for natural flocks of birds,"
Proc. Natl Acad. Sci USA {\bf 109}, 4786 (2012)
[arXiv:1107.0604 [physics.bio-ph]].

\bibitem{dynamical crit review}
A.~Roli, M.~Villani, A.~Filisetti and R.~Serra,
``Dynamical criticality: overview and open questions,"
J. of Systems Sci and Complexity, {\bf 31}, 647 (2018).

\bibitem{Kauffman}
S.~A.~Kauffman, 
``The Origins of Order: Self-Organization and Selection in Evolution," 
Oxford University Press, Oxford (1993).

\bibitem{wolfram}
S.~Wolfram,
``Universality and complexity in cellular automata,"
Physica D {\bf 10}, 1 (1984).

\bibitem{edge of chaos 1}
N.~H.~Packard, 
``Adaptation toward the edge of chaos," 
Dynamic Patterns in Complex Systems {\bf 212}, 293 (1988).

\bibitem{edge of chaos 2}
C.~G.~Langton,
``Computation at the edge of chaos: Phase transitions and emergent computation," 
Physica D {\bf 42}, 12 (1990).

\bibitem{edge of chaos 3}
J.~P.~Crutchfield and K.~Young,
``Inferring statistical complexity,"
Phys. Rev. Lett. {\bf 63}, 105 (1989).

\bibitem{edge of chaos 4}
J.~P.~Crutchfield and K.~Young,
``Computation at the onset of chaos,"
In W. H. Zurek, editor, {\it Complexity, Entropy, and the Physics of Information}, p. 223–269, Addison-Wesley (1990).

\bibitem{edge of chaos 5}
M.~Mitchell, P.~Hraber and J.~P.~Crutchfield,
``Revisiting the Edge of Chaos: Evolving Cellular Automata to Perform Computations,"
Complex Systems {\bf 7}, 89 (1993)
[arXiv:adap-org/9303003].
 
\bibitem{RNNchaos1}
N.~Bertschinger and T.~Natschl\"ager,
``Real-time computation at the edge of chaos in recurrent neural networks,"
Neural Computation {\bf 16}, 1413 (2004).

\bibitem{RNNchaos2}
J.~Boedecker, O.~Obst O, J.~T.~Lizier, N.~M.~Mayer and M.~Asada,
``Information processing in echo state networks at the edge of chaos,"
Theory in Biosciences {\bf 131}, 205 (2012).

\bibitem{RNNchaos3}
F.~Matzner,
``Neuroevolution on the Edge of Chaos,"
Proc. of the Genetic and Evolutionary Computation Conference, 465 (2017)
[arXiv:1706.01330 [cs.NE]].

\bibitem{DNNpowerlaw}
C.~H.~Martin and M.~W.~Mahoney,
``Implicit Self-Regularization in Deep Neural Networks: Evidence from Random Matrix Theory and Implications for Learning,"
arXiv:1810.01075 [cs.LG].

\bibitem{RMT}
D.~Sornette,
``Critical phenomena in natural sciences: chaos, fractals, self-organization and disorder: concepts and tools,"
Springer-Verlag, Berlin, 2006.

%\cite{Gunion:1987qv}
\bibitem{Gunion:1987qv} 
  J.~F.~Gunion, H.~E.~Haber and M.~Sher,
  ``Charge / Color Breaking Minima and a-Parameter Bounds in Supersymmetric Models,''
  Nucl.\ Phys.\ B {\bf 306}, 1 (1988).
%  doi:10.1016/0550-3213(88)90168-X

\bibitem{Giudice:2011cg}
 G.~F.~Giudice and A.~Strumia,
  ``Probing High-Scale and Split Supersymmetry with Higgs Mass Measurements,''
  Nucl.\ Phys.\ B {\bf 858}, 63 (2012)
 % doi:10.1016/j.nuclphysb.2012.01.001
  [arXiv:1108.6077 [hep-ph]].
  
%\cite{Hertzberg:2012zc}
\bibitem{Hertzberg:2012zc} 
  M.~P.~Hertzberg,
  ``A Correlation Between the Higgs Mass and Dark Matter,''
  Adv.\ High Energy Phys.\  {\bf 2017}, 6295927 (2017)
%  doi:10.1155/2017/6295927
  [arXiv:1210.3624 [hep-ph]].


\end{thebibliography}
\end{document}